\documentclass[fleqn,usenatbib]{mnras}
\usepackage[T1]{fontenc}
\usepackage{amsmath, amssymb}
\usepackage{mathtools} %
\usepackage{newtxtext,newtxmath}
\DeclareRobustCommand{\VAN}[3]{#2}
\let\VANthebibliography\thebibliography
\def\thebibliography{\DeclareRobustCommand{\VAN}[3]{##3}\VANthebibliography}
\usepackage{graphicx}
\usepackage[UKenglish]{babel} %
\usepackage{nicefrac} %
\usepackage[style=UKenglish]{csquotes} %
\usepackage[normalem]{ulem} %
\usepackage[usenames,dvipsnames]{xcolor} %
\usepackage{xspace} %
\usepackage{wasysym}
\usepackage{mathrsfs} %
\usepackage{placeins} %
\usepackage{natbib}
\usepackage{subcaption} %
\usepackage{hyperref} %

\makeatletter
\NewDocumentCommand\TODO{m}{
    \@latex@warning{TODO: #1}
    \textcolor{red}{TODO: #1}
}
\makeatother

\newcommand{\twocm}[0]{21\,cm\xspace} %
\newcommand{\twocmtitle}{\texorpdfstring{21\,cm}{21 cm}\xspace} %

\newcommand{\rr}[1]{\color{black}{#1}\color{black}\xspace}

\graphicspath{{./figures/}}

\let\jnlstyle=\rm
\def\jref#1{{\jnlstyle#1}}
\def\apj{\jref{ApJ}}                 %
\def\apjl{\jref{ApJ}}                %
\def\aap{\jref{A\&A}}                %
\def\jcap{\jref{J. Cosmology Astropart. Phys.}}
\def\mnras{\jref{MNRAS}}             %
\def\prd{\jref{Phys.~Rev.~D}}        %
\def\pasa{\jref{PASA}}               %
\def\pasp{\jref{PASP}}               %
\def\nat{\jref{Nature}}              %
\def\physrep{\jref{Phys.~Rep.}}   %

\title[FlexKnot \& Gaussian Process for \twocmtitle global signal]{FlexKnot and Gaussian Process for \twocmtitle global signal analysis and foreground separation}
\author[S. Heimersheim et al.]{
Stefan Heimersheim,$^{1}$\thanks{E-mail: heimersheim@ast.cam.ac.uk}
Leiv R\o nneberg,$^{2,3}$
Henry Linton$^{4,1}$
Filippo Pagani,$^{2}$
and Anastasia Fialkov$^{1,5}$
\\
$^{1}$Institute of Astronomy, University of Cambridge, Madingley Road, Cambridge CB3 0HA, UK\\
$^{2}$MRC Biostatistics Unit, University of Cambridge, Cambridge CB2 0SR, UK\\
$^{3}$Oslo Centre for Biostatistics and Epidemiology, University of Oslo, Oslo, Norway\\
$^{4}$Physics Department, Blackett Lab, Imperial College, Prince Consort Road, London SW7 2AZ, UK \\
$^{5}$Kavli Institute for Cosmology, Madingley Road, Cambridge CB3 0HA, UK
}
\date{Accepted 2023 December 13. Received 2023 December 08; in original form 2023 September 07}
\pubyear{2023}

\begin{document}
\label{firstpage}
\pagerange{\pageref{firstpage}--\pageref{lastpage}}
\maketitle

\begin{abstract}
	The cosmological \twocm signal is one of the most promising avenues to study the Epoch of Reionization. One class of experiments aiming to detect this signal is global signal experiments measuring the sky-averaged \twocm brightness temperature as a function of frequency. A crucial step in the interpretation and analysis of such measurements is separating foreground contributions from the remainder of the signal, requiring accurate models for both components. Current models for the signal (non-foreground) component, which may contain cosmological and systematic contributions, are incomplete and unable to capture the full signal.
	We propose two new methods for extracting this component from the data: Firstly, we employ a foreground-orthogonal Gaussian Process to extract the part of the signal that cannot be explained by the foregrounds.
	Secondly, we use a FlexKnot parameterization to model the full signal component in a free-form manner, not assuming any particular shape or functional form. This method uses Bayesian model selection to find the simplest signal that can explain the data.
	We test our methods on both, synthetic data and publicly available EDGES low-band data. We find that the Gaussian Process can clearly capture the foreground-orthogonal signal component of both data sets.
	The FlexKnot method correctly recovers the full shape of the input signal used in the synthetic data and yields a multi-modal distribution
	of different signal shapes that can explain the EDGES observations.
\end{abstract}


\section{Introduction}
\label{global:sec:intro}
\twocm cosmology is the endeavour to use the hyperfine transition
of neutral hydrogen to probe the Universe at
high redshifts ($5 \lesssim z \lesssim 30$), from the Dark Ages,
when the Universe was neutral via Cosmic Dawn when the first stars formed, to the Epoch of Reionization
(EoR), when early galaxies ionized the intergalactic
medium (IGM). This is one of the last unexplored
frontiers of cosmology, and measurements of the
\twocm signal of neutral hydrogen will allow us
to study the properties of the IGM and the first
stars and galaxies \citep{Madau97, Furlanetto2006c, Pritchard2012, Barkana2016}.

The \twocm signal is the emission or absorption line
corresponding to the
hyperfine transition of neutral hydrogen. When the spin
temperature of the hydrogen atoms is higher than the
background radiation temperature, the atoms emit, creating
a peak in the differential brightness temperature.
When it is lower,
the hydrogen atoms absorb the background radiation
causing a trough in the spectrum. We generally expect
a trough during Cosmic Dawn and potentially a peak
towards lower redshifts when the IGM is heated above
the background temperature.
We can observe this signal as a distortion in a smooth
radio background (typically assumed to be the
Cosmic Microwave Background, CMB), at a frequency
of $\nu_{21} = 1420.4\,{\rm MHz} / (1+z)$ depending
on the redshift $z$ of the corresponding hydrogen cloud.

The two main types of \twocm experiments are global
signal experiments, which measure the sky-averaged
\twocm signal, and interferometers, which
measure the spatial fluctuations of the signal.
Global signal experiments include
EDGES \citep{Bowman2008},
SARAS \citep{Patra2013},
SCI-HI \citep{Voytek2014},
BIGHORNS \citep{Sokolowski2015},
LEDA \citep{LEDA,Garsden_2021},
PRIZM \citep{PRIZM},
MIST \citep{MIST_2023arXiv230902996M},
and
REACH \citep{REACH}.
Of particular interest are EDGES, which has claimed
a detection of the \twocm signal \citep{Bowman2018},
and SARAS 3, which reported a non-detection
\citep{Singh2022} of the signal claimed by EDGES.
Future experiments (e.g. REACH and MIST) are expected to
provide independent global signal measurements to
answer this question.
Measurements of fluctuations in the \twocm signal have been
performed by several interferometers. While no detection
has been reported so far, these measurements have set
upper limits on the \twocm power spectrum.
The strongest limits so far have been set by the HERA
telescope \citep{h1cidr2limits,hera2023}; other
interferometers include
LOFAR \citep{Patil2017, Mertens2020},
MWA \citep{Dillon2014,Trott2020},
NenuFAR \citep{NenuFAR},
PAPER \citep{Kolopanis2019},
GMRT \citep{Paciga2013}.
Although fluctuations in the cosmological signal contain unique
information about the first sources of light and the underlying
cosmology, in this paper we choose to focus on the global \twocm
signal measurement.

The \twocm global signal is challenging to observe.
Not only does a measurement require sufficient sensitivity \citep[the
signal reaches a contrast of up to
$280\,\mathrm{mK}$,][under the assumption that the background is the CMB, ]{Cohen2017},
but it is also contaminated by
astrophysical foreground sources \citep{Chandrasekhar1960book,Rogers2015},
radio frequency interference \citep[RFI,][]{Rogers2005},
ionospheric effects \citep[][]{Vedantham2014,Datta20142016iono,Shen2021,Shen2022},
and instrumental effects such as beam chromaticity \citep[e.g.][]{Monsalve2017, Anstey2021}
or analogue receiver properties \citep{Nambissan2021SARAS3receiver, REACH}.
While instrumental challenges can be addressed by careful design
and calibration \citep{Monsalve2017a}, the unknown astrophysical foreground
sources have to be constrained jointly with the signal.
Thus, we fit the measured sky temperature as a sum of the foreground
and signal components 
$T_{\rm sky}(\nu) = T_{\rm fg}(\nu) + T_{\rm signal}(\nu)$, following
\citet{Bowman2018}.

The claimed detection of the \twocm signal by \citet{Bowman2018} showed
a much stronger absorption trough than expected from physical models
that assume the CMB as the background radiation and standard astrophysical
heating and cooling processes \citep{Cohen2017,Reis2021}.
This has sparked discussion about the
choice of foreground model \citep{Hills2018,Singh2019,Bevins2021maxsmooth},
the question of instrumental calibration \citep{Sims2020}, and the possibility
of non-standard cosmological signals \citep{Barkana2018,Ewall_Wice_2018,Feng:2018,Munoz:2018pzp,Slatyer:2018aqg,Fialkov2019,mirocha19}.
However, all previous work either assumed the signal to have a
particular shape that does not necessarily correspond to the actual
signal \citep[including the original work by][]{Bowman2018}, or used
physical models, typically based on simulations \citep[such as][]{
	Furlanetto2004,
	Iliev06_sim, %
	McQuinn07, %
	Mesinger2011, %
	Fialkov2014a, %
	Ross17, %
	Semelin17} %
that are limited by our understanding of the underlying astrophysics \citep{Majumdar2014}.
This mismatch between the signal model and the true signal will affect the quality of the foreground separation and prevent the extraction of the correct signal from the data.

In this work, we propose a new way to model the signal component
that is agnostic to the precise form of the signal, allowing for
an unexpected cosmological signal shape or an undetected systematic
contribution. From this point on, we will refer to any non-foreground part of the
measurement as the \textit{signal} component here, and do not address
the question of separating the \textit{cosmological} signal
from systematics. This question is considered in the forthcoming
work by \citet{Shen2023arXiv231114537S}.
Concretely we propose two techniques to describe
this signal component: A non-parametric
Gaussian Process, and a model-independent
FlexKnot parameterization. These free-form approaches
allow us to find out what (not necessarily cosmological)
signal is hidden in the data. This signal can then be
interpreted, either to learn about systematics in the data
or about astrophysical processes. The latter will however
require physical models to connect the observed signal
properties to physical properties, re-introducing model-dependence.

Gaussian Processes \citep[GPs,][]{GPbook} are a standard tool for modelling functions in a non-parametric fashion. Taken as a prior over smooth functions, GP regression enables Bayesian inference over a regression function with an unknown parametric form and has successfully been used in a wide array of applications \citep{Almosallam2016,Ghosh2020,Li2021,Mertens2023}\footnote{Note that concurrent work by \citet{Mertens2023} uses GPs to model the \twocm signal based on semi-numerical simulations, while our goal is to achieve modelling of the signal \textit{without} relying on simulations.}. Utilizing a GP to model the unknown signal allows us to place a prior distribution over the signal without assuming a known shape. Gaussian Processes
have a small number of parameters specifying the mean and covariance but these
only encode priors on the signal smoothness and strength and do not
impose any functional form on the signal.
However, since GPs are very flexible function approximators, they need to be sufficiently constrained to avoid confounding with the foreground model. 
We make use of quite a harsh constraint to obtain this, relying on orthogonalizing the GP with respect to the foreground fit, but other options are possible as well.

In this work, we utilize the concept of a
foreground-orthogonal basis, that has been
previously considered in the literature \citep{Liu2012,Vedantham2014,Switzer2014,
Tauscher2018,Tauscher2020,Rapetti2020,LiuShaw2020,Bassett2021}
We say a signal function is orthogonal
to the foregrounds if no amount of foreground
terms can reduce the
norm of the signal. If we consider the
foreground terms and signal as vectors containing
their value at each of the observed frequencies then this orthogonality is equivalent
to the signal vector being orthogonal (zero dot product) to all foreground term vectors.
This concept is particularly useful here because the foreground-orthogonal component of a measurement is unambiguous (although still dependent on the foreground model) and has to be fully contained within the signal component (which could be
of cosmological origin or caused by systematics).

In addition to the Gaussian Process, we explore a second method to fit
the signal in a model-independent way.
The FlexKnot parameterization is based on a free-form method that has been
\rr{introduced by \citet{2009MNRAS.400.1075B} for modeling the primordial
power spectrum. It has been used in the
CMB community for several years, constraining the primordial
power spectrum and dark energy equation of state \citep{Vazquez2012,2012JCAP...09..020V,2013JCAP...08..001V,2014JCAP...08..053A,2014JCAP...08..052A,2016MNRAS.455.2461H,2016A&A...594A..20P,2017MNRAS.466..369H,2018JCAP...04..016F,2019PhRvD.100j3511H,2020A&A...641A...1P,2020A&A...641A..10P,2023EPJC...83..251E}. It was also used to model the pressure profile of galaxy clusters \citep{2018MNRAS.481.3853O}.
\citet{Millea2018} adapted the method for parameterizing the reionization history, giving it the
name FlexKnot.}
More recently the use of FlexKnot has been extended
by \citet{Heimersheim2022FRB}
to explore reionization history constraints from hypothetical high-redshift
Fast Radio Bursts.
Their implementation forms the basis of the implementation adopted in this work.

The FlexKnot method differs from the Gaussian Process in that it is
a parametric method, and allows for a variable amount of complexity.
The parameterization works by specifying a set of interpolation
points (knots) as $x$ and $y$ coordinates (the parameters)
that allow us to describe the signal function by interpolating between these knots.
The number of interpolation knots is itself also a parameter, leading
to the key difference between the Gaussian Process and FlexKnot methods:
We can compare interpolations of different complexity levels and use Bayesian model selection to find the \textit{simplest} signal that can explain the observations. At the same time, we still obtain a free-form signal fit that is not limited to a particular shape.

With this paper, we introduce the concept of a free-form model
for the global \twocm signal. We highlight what the Gaussian Process
and FlexKnot methods can
and cannot do, and demonstrate their use on both synthetic data,
and the EDGES low-band measurements from \citet{Bowman2018}. 
We describe both methods in section \ref{global:sec:method},
present the results for both cases in section \ref{global:sec:results},
discuss advantages and limitations in section \ref{global:sec:discussion},
and give our conclusions in section \ref{global:sec:conclusion}.

\section{Method}
\label{global:sec:method}
The quantity observed by a global \twocm signal experiment is the
sky-averaged brightness temperature as a function of frequency,
$T_{\rm sky}(\nu)$. The actual measurement is a spectral intensity
but it is typically expressed as the corresponding blackbody temperature.
The sky temperature is dominated by contributions from foregrounds
but contains a small cosmological signal that we are interested in.
Additionally, the measured temperature may contain systematic
contributions. In this work, we focus on separating the foregrounds from the remaining signal, $T_{\rm signal}(\nu)$,
and will not distinguish between cosmological
and systematic contributions \citep[this is addressed in the forthcoming
work by][]{Shen2023arXiv231114537S}.
Thus, for our purposes, the sky-averaged
observed brightness temperature is
\begin{equation}
	T_{\rm sky}(\nu) = T_{\rm fg}(\nu) + T_{\rm signal}(\nu)
\end{equation}
where $T_{\rm fg}(\nu)$ is the foreground component and $T_{\rm signal}(\nu)$
contains the remaining contributions (including the cosmological signal and possible systematics).

In section \ref{global:subsec:method_foreground} we briefly describe the foreground model we adopt from \citet{Bowman2018}. Section \ref{global:subsec:method_data}
describes the synthetic data we generate to validate our methods and section
\ref{global:subsec:method_likelihood} summarizes the likelihood function adopted from
\citet{Bowman2018}. Finally, we give a detailed description of the Gaussian Process and FlexKnot methods in sections \ref{global:subsec:method_GP} and \ref{global:subsec:method_FK} respectively.

\subsection{Foreground model}
\label{global:subsec:method_foreground}
The core advances discussed in this work concern the modelling of the
signal component, hence we only briefly describe the foreground model
and use a simple model derived in \citet{Bowman2018}. However,
our signal modelling approach can be used with any parameterized foreground
model \citep[e.g. the alternatives considered in][]{Hills2018}.

For concreteness, we assume the following parameterized foreground
model. It describes the foreground contribution to the sky temperature
as a sum of polynomial terms
\begin{equation}
	T_{\rm fg}(\nu) = \sum_{n=0}^{4} a_n \underbrace{\left(\frac{\nu}{\nu_c}\right)^{n-2.5}}_{f_n(\nu)}
	\label{global:eq:foreground_polynomial}
\end{equation}
where $a_n$ are free parameters and $\nu_c$ is set to $75\,\mathrm{MHz}$.
We note that this model corresponds to equation (2) rather than (1)
of \citet{Bowman2018} but \citet{Bowman2018} themselves report consistent
results for both models.
We assume wide and nearly uninformative priors on the foreground parameters, as we discuss in sections \ref{global:subsubsec:method_GP_priors} and \ref{global:subsubsec:method_FK_priors} for each method.

\subsection{Synthetic data}
\label{global:subsec:method_data}
We generate a synthetic data set with a known ground truth
to validate our methods. We use the foreground model from
equation \eqref{global:eq:foreground_polynomial}, and a
signal component corresponding to a flattened
Gaussian \citep[from][]{Bowman2018}
\begin{equation}
\begin{gathered}
	T_{\rm signal} = -A \left(\frac{1-\exp^{\left(-\tau \mathrm e^B\right)}}{1-\mathrm e^{-\tau}}\right)
	\\
	\text{with } B = \frac{4(\nu-\nu_0)^2}{w^2} \log\left[\frac{-1}{\tau}\log\left(
		\frac{1+\mathrm e^{-\tau}}{2}\right)\right]
	\label{global:eq:flattened_gaussian}
\end{gathered}
\end{equation}
with amplitude $A$, central frequency $\nu_0$, full-width at half-maximum $w$,
and flattening factor $\tau$. Finally, we add Gaussian noise with a
standard deviation of $25\,\mathrm{mK}$
to emulate the thermal noise level of the EDGES experiment \citep{Bowman2018}.
The synthetic data are generated with foreground parameters
$(a_1, a_2, a_3, a_4, a_5) = (1570, 620, -1000, 700, -175)$
and signal parameters
$(A, \nu_0, w, \tau) = (500\,\mathrm{mK}, 78\,\mathrm{MHz}, 19\,\mathrm{MHz}, 7)$;
both sets were chosen to be similar to those found in \citet{Bowman2018}.\footnote{Note that we do not expect our synthetic data to exactly match the EDGES 
observations. The reasons for this are (i) that we picked a set of plausible 
parameters but not exactly the best-fitting parameters from EDGES, (ii) that
we use a slightly different foreground model, and (iii) that the EDGES
observations may contain a signal that is different from the flattened
Gaussian we assumed for the synthetic data.}

Note that we use equation \eqref{global:eq:flattened_gaussian},
not only to generate the synthetic data (with the parameter values
mentioned above) but separately also to fit the EDGES
low-band data. In the latter case, we \textit{fit} the
parameters of equation \eqref{global:eq:flattened_gaussian} in
order to reproduce the method of \citet{Bowman2018} as a comparison
to our methods. We emphasize that these are two separate uses of
equation \eqref{global:eq:flattened_gaussian}, with different
signal parameters and different foregrounds.

\subsection{Likelihood function}
\label{global:subsec:method_likelihood}
To compare a given model $\mathbf{m}=\mathbf{T}_{\rm fg}+\mathbf{T}_{\rm signal}$
(a vector composed of temperature values at the observed frequencies)
to the measured data $\mathbf{d}$ (vector of measurements at every frequency)
we use a Gaussian likelihood, following \citet{Bowman2018}.
The likelihood accounts for Gaussian thermal noise with a free parameter
$\sigma_{\rm noise}$ that determines the standard deviation:
\begin{gather}
	\label{global:eq:likelihood_simple}
	P(\mathbf{d} \mid \mathbf{m}) =
	\frac{1}{\sqrt{2\pi}\,\sigma_{\rm noise}}
	\exp{\left(-
	\frac{\left(\mathbf{d}-\mathbf{m}\right)^2}{2\sigma_{\rm noise}^2}
	\right)}.
\end{gather}
We use this likelihood for both our synthetic data and the EDGES data.
Since the EDGES low-band data are measured in 123 frequency channels
\citep{Bowman2018} $\nu_1,\ldots,\nu_{123}$, the data as well as the signal
and foregrounds will be 123-dimensional temperature vectors corresponding
to those frequencies. 

\subsection{Gaussian Process signal model}
\label{global:subsec:method_GP}
We now proceed with introducing the methods to represent the signal
component, starting with the Gaussian Process approach. GPs
are a common way to model smoothly varying functions. A GP is a stochastic process,
i.e. a collection of random variables, defined such that the joint distribution of any finite sub-collection is a multivariate normal. 
Specifically, we consider the continuous Gaussian Process
$\mathcal{GP}(\mu(\nu),\kappa({\nu,\nu'}))$ to describe the signal
at a frequency $\nu$, 
where $\mu(\nu)$ is the mean function calculated at frequency $\nu$, and $\kappa(\nu,\nu')$ is the covariance function calculated between two frequencies $\nu$ and $\nu'$.
We generate the GP signal vector as a finite realization of
a Gaussian Process. It follows a multivariate normal distribution
$\left[T_{\rm GP}(\nu_1),\ldots,T_{\rm GP}(\nu_{n})\right]^T \sim \mathcal{N}(\boldsymbol{\mu},K)$
where $\boldsymbol{\mu}=[\mu(\nu_1),\ldots,\mu(\nu_n)]^T$ denotes the mean vector constructed by evaluating the mean function at each input location.
Correspondingly, the covariance matrix $K$ is defined such that $K_{ij}=\kappa(\nu_i,\nu_j)$ for all frequency pairs $(\nu_i, \nu_j)$. The GP is fully specified by its mean and covariance functions. Of particular importance is the covariance function, sometimes denoted as the \textit{kernel} of the GP, which controls many features of the resulting function such as the level of smoothness (correlation) across
the input locations.

We choose the mean function to be zero, $\boldsymbol{\mu}=\mathbf{0}$,
to not bias the signal towards any particular shape.
As the covariance matrix, we choose a simple squared exponential kernel
\begin{equation}
	\label{eq:GP_kernel}
	K_{ij} = \kappa(\nu_i, \nu_j) = \sigma_f^2 \exp {\left(-\frac{(\nu_i - \nu_j)^2}{2\ell^2}\right)}
\end{equation}
with amplitude $\sigma_f^2$ and length scale $\ell$. The
amplitude controls how far the GP samples are allowed to move
away from the mean, and the length scale controls the overall
correlation length of the samples across frequency space.

In addition to this smooth variation, we allow for thermal noise as described
in section \ref{global:subsec:method_likelihood}.
Taking equation \eqref{global:eq:likelihood_simple} with
$\mathbf{m}=\mathbf{T}_{\rm fg}+\mathbf{T}_{\rm GP}$ yields the
corresponding likelihood for observational data $\mathbf{d}$.
In order to draw samples from the Gaussian Process we 
use a formulation that explicitly includes the noise
term $\epsilon_i$, i.e. we draw expected measurements at
frequency $\nu_i$ from the model
$T_{\rm fg}(\nu_i) + T_{\mathrm{GP}}(\nu_i) + \epsilon_i$
with $\epsilon_i$ being a Gaussian random variable with zero mean and
variance $\sigma_{\rm noise}^2$.

\subsubsection{Orthogonal Gaussian Process}
\label{global:subsubsec:method_GP_orthogonal}
\begin{figure}
	\centering
	\includegraphics[width=\linewidth]{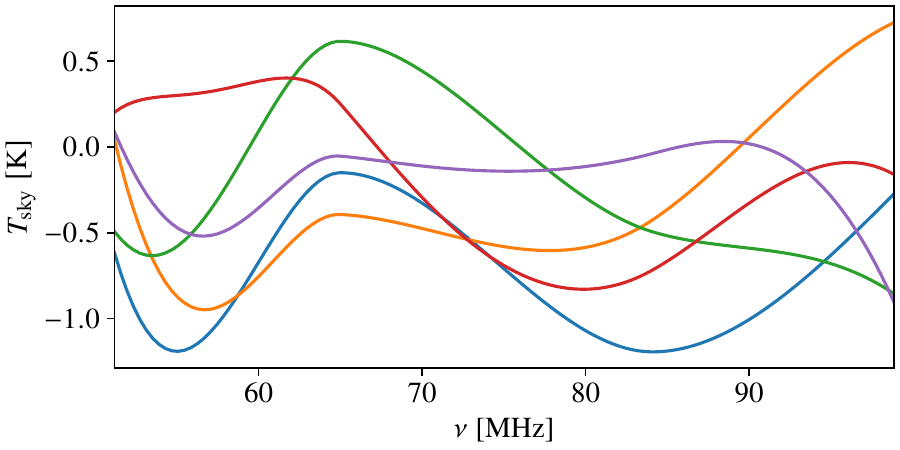}
	\caption{Five different signals that are equivalent up to
	their foreground-proportional terms. Each of these will
	have the same likelihood (for different values of the
	foreground parameters), thus making the signals equally
	preferred by the data.}
	\label{fig:illustration_fgvariation}
\end{figure}

\rr{It should be noted that combining the parametric model with the GP with no further constraints yields a solution characterised by degeneracy between estimates of the foregrounds, and signal. To illustrate this point, consider the signals in Figure \ref{fig:illustration_fgvariation}.}
All the demonstrated curves contain the same cosmological signal
but differ in the foreground component, and are thus equally preferred
by the data (as any \enquote{foreground-shaped}
addition can be compensated by slightly different foreground parameters).
\rr{This leads to confounding \citep{hodges2010}, as the fit is jointly appropriate, but there
is no incentive for the GP to separate the signal from the foregrounds.
This degeneracy leads to large uncertainties on both the GP signal
and the foreground parameters.}

\rr{To control the degeneracy between the GP and the foreground model, the GP must be penalized away from functional forms that are similar to the foreground. One way to do this is through penalization of the kernel hyperparameters $\ell$ and $\sigma_f$, for example, a prior choice for $\sigma_f$ that encourages small deviations from the prior mean at zero. That choice however could also negatively affect the result, e.g. by discouraging a better-fitting signal requiring larger deviations from the prior mean. 

Another way to avoid the degeneracy, and which will be our main method pursued in the paper, is to only fit the foreground-orthogonal component
of the signal, rather than the full signal.} The foreground-orthogonal part contains
less information than the full signal because there exist multiple signals
that have the same foreground-orthogonal component (e.g. those shown in
Figure \ref{fig:illustration_fgvariation}). However, these are exactly
the kind of signals that cannot be distinguished by the likelihood.
Thus we fit the foreground-orthogonal component which presents
exactly the information that can be extracted from the data alone.
In section \ref{global:subsec:result_GP} we will see that the Gaussian Process can very accurately estimate this component of the signal.

\begin{figure}
	\centering
	\includegraphics[width=\linewidth]{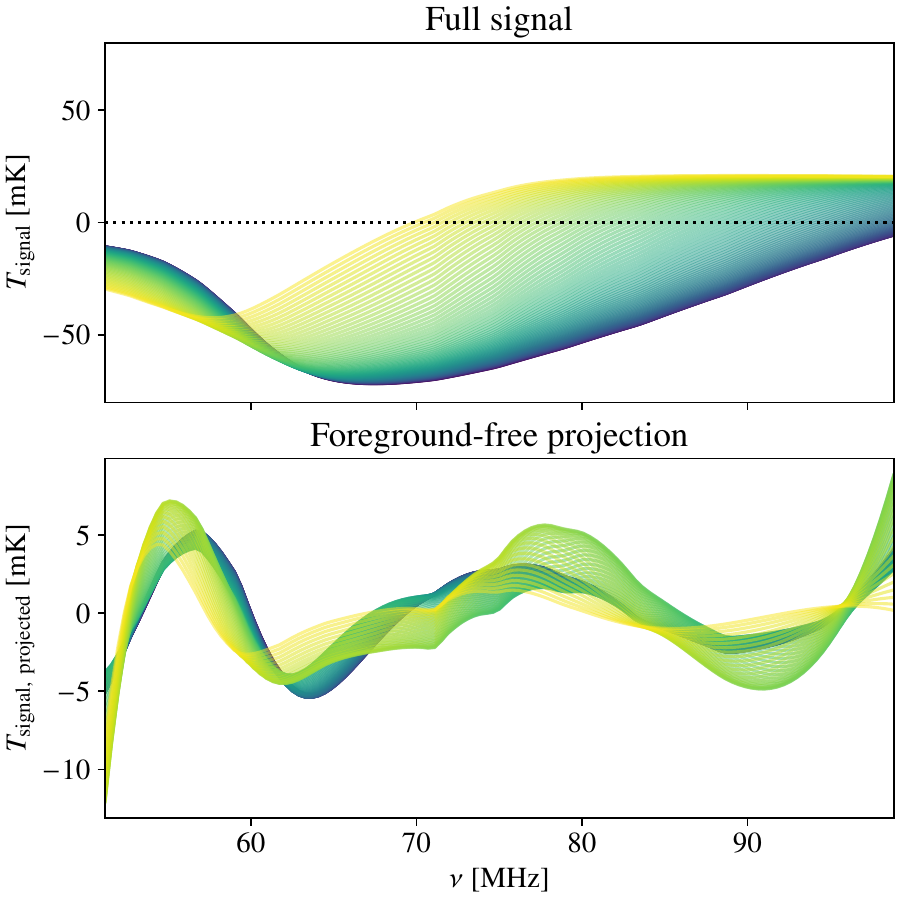}
	\caption[placeholder short caption]{Demonstration of the foreground-orthogonal projection
	with a range of simulated cosmological signals. The particular signals
	are chosen for illustrative purposes, the different lines correspond
	to different settings of the X-ray production in early galaxies
	($f_X$ from $10^{-1}$, blue, to $10^3$, yellow, see
	\citet{Fialkov2014a,Fialkov2014b,Fialkov2014c} for details\footnotemark).
	The top panel shows $T_{\rm signal}$ with the recognizable shape,
	but the bottom panel shows the foreground-orthogonal component
	$T_{\rm signal,\ projected}$ of these signals (note the $y$ axis scale).
	We want to stress that we show these cosmological signals without any
	systematics and only for illustrative purposes,
	while $T_{\rm signal}$ in our analysis may contain both cosmological and systematic contributions.
	}
	\label{global:fig:illustration_astrosignals}
\end{figure}
\footnotetext{The remaining simulation parameters are set to
$R_{\rm mfp}=40\,\mathrm{Mpc}$,
$\zeta=15$,
$V_{c}=15\,\mathrm{km/s}$,
$f_{*,\mathrm{II}}=0.02$,
$f_{*,\mathrm{III}}=0.01$,
$f_r=1$,
and the intermediate stellar initial mass function
from \citet{Gessey-Jones2022}.}

The foreground-orthogonal signal fit obtained from the GP is 
generically useful for signal comparison purposes. In particular,
it can be used to compare the data to (e.g. physical or systematic) theory
signals by simply comparing their foreground-orthogonal components.
This can be done by applying the foreground-orthogonal projection
to the theory signal and then comparing the resulting
foreground-orthogonal theory to the foreground-orthogonal GP fit.
We illustrate the projection in Figure \ref{global:fig:illustration_astrosignals},
showing full theory signals (top panel) and their foreground-orthogonal
components (bottom panel).

Comparing foreground-orthogonal projections is sufficient and lossless
(under a Gaussian likelihood as in equation \eqref{global:eq:likelihood_simple},
and an unrestricted foreground model as adopted here).
If the foreground-orthogonal theory signal is close to the foreground-orthogonal
component of the data, this implies that there exists a set of foreground parameters
for which the theory model plus a foreground term are equally close to
the full data. We give a mathematical derivation of this in the paragraph on
analytical foreground marginalization in section \ref{global:subsubsec:method_ortho_fg}.

In the future, it might be practical for signal interpretation
purposes to consider the foreground-orthogonal components,
rather than analyze the full signals. Projecting an ensemble
of physical signals would enable the interpretation of the
foreground-orthogonal component in terms of astrophysical properties,
just as it is done now with full signals. We discuss this option
further in section \ref{global:sec:discussion}.

To create estimate the foreground-orthogonal component we construct
a Gaussian Process that is always orthogonal to the foregrounds,
allowing us to fit the parametric foregrounds and non-parametric GP
signal simultaneously without degeneracy.
While this is generally difficult to do for GPs across the entire input space
 \citep[see e.g.][]{plumlee_orthogonal_2017}, it can easily be achieved at any finite
set of locations by \textit{conditioning by kriging} \citep{Rue2005}. These processes
have been utilized in spatial statistics under the name
\emph{restricted spatial regression} 
\citep[RSR,][]{hanks_restricted_2015,khan_restricted_2022}.

We can achieve this by first drawing $\mathbf{T}_{\rm GP}$ from the unconstrained
multivariate Normal distribution ${\mathcal{N}(\mathbf{0},K)}$ and then
applying the transformation
${\mathbf{T}_{\rm GP, proj} = (\mathrm{Id} - P_F)\,\mathbf{T}_{\rm GP}}$
using the projection matrix $P_F = KF^T(FKF^T)^{-1}F$ with the $5\times123$-dimensional design matrix $F_{ni}=f_n(\nu_i)$ describing the
5 independent foreground modes from equation
\eqref{global:eq:foreground_polynomial}. \rr{This can be derived by first considering the joint distribution of the GP $\mathbf{T}_{\rm GP}$ and a linear transformation of the GP $F\mathbf{T}_{\rm GP}$ -- which will be multivariate normal. Then the conditional distribution of $\mathbf{T}_{\rm GP}\vert F\mathbf{T}_{\rm GP} = \mathbf{0}$ is available through standard formulas, enforcing the orthogonality constraint.} 
To efficiently sample from the posterior distribution, we use the standard trick of marginalizing out the GP from the joint distribution to obtain the following distribution of our data:
\begin{equation}\label{eq:GPsampled}
	T_\mathrm{obs} \sim \mathcal{N}\left(T_{\rm fg}, (\mathrm{Id} - P_F) K
	(\mathrm{Id}  - P_F)^T + \sigma_{\rm noise}^2 \mathrm{Id} \right).
\end{equation}
That is, we assume our observed data is drawn from a normal distribution centred around the foreground model, with spatial covariance defined by the orthogonal GP. \rr{Conditional on the foreground, kernel hyperparameters, and noise parameter $\sigma^2$, the posterior distribution of the 21cm signal is available in closed form, and is simply a multivariate normal $\mathcal{N}(\tilde{K}(\tilde{K}+\sigma^2\mathrm{Id})^{-1}(T_\mathrm{obs}-T_{\rm fg}),\tilde{K}-\tilde{K}(\tilde{K}+\sigma^2\mathrm{Id})^{-1}\tilde{K})$ where $\tilde{K}=(\mathrm{Id} - P_F) K
	(\mathrm{Id}  - P_F)^T$, following the standard formulas of Gaussian Process regression. We sample from this distribution at each step of an MCMC chain to recover the posterior distribution of $\mathbf{T}_{\rm GP, proj}$ itself.}

\subsubsection{Gaussian Process priors}
\label{global:subsubsec:method_GP_priors}
We expect the variation of the GP to be of the order of $\pm 1\,\mathrm K$ so
we set the prior on $\sigma_f$ to be the exponential distribution
$\pi(\sigma) = \exp(-\sigma\,\mathrm{K}^{-1})$. This standard choice allows
deviation from the mean with a preference for smaller fluctuations (we want this preference to not introduce artificial fluctuations due to the prior).
If one is expecting signals much larger than $1\,\mathrm{K}$
\citep[e.g. in cases with additional radio backgrounds present or non-standard thermal histories][]{Barkana2018,Ewall_Wice_2018,Feng:2018,Munoz:2018pzp,Slatyer:2018aqg,Fialkov2019,mirocha19} this choice can be adjusted.
As for the smoothness we are working in the frequency range
$[50\,\mathrm{MHz}, 100\,\mathrm{MHz}]$ and set the prior on $\ell$ to
be an inverse-gamma distribution with shape $\alpha=5$
and scale $\beta=75\,\mathrm{MHz}$. This is a fairly standard choice where we
penalize the model away from exceedingly small or large length scales, but the
GP is not very sensitive to these choices due to the orthogonality constraint.

The remaining priors are set as follows.
For the foreground component, we place uninformative priors on the parameters of the foreground model using the recommended QR reparametrization trick as described in the Stan manual \citep{stanManual}. The prior for each scaled and transformed parameter is given by a wide and uninformative Normal distribution $\mathcal{N}(0,10^6)$.
For the noise $\sigma_{\rm noise}$ we assume a half-Normal distribution
value\ $\mathcal{N}^+(0\,{\rm K},1\,{\rm K})$ to allow for a wide range
covering all reasonable values. \rr{We do full MCMC for all parameters in the model, and implement the orthogonal GP model in Stan  \citep[RStan version 2.26,][]{rstan,stancore} using the No-U-Turn-Sampler algorithm} \citep{Hoffman2014}. 
The results of estimating the Gaussian Process are shown in section \ref{global:subsec:result_GP}.
\rr{Note that we also did fit a non-orthogonalised GP to the data, using the same prior distributions for the hyperparameters as above, but found that it was difficult to avoid degeneracy with the foreground through prior distributions only.}

\subsection{FlexKnot signal model}
\label{global:subsec:method_FK}

The Gaussian Process fulfils one of our goals: it allows us to extract
information about the signal model without assuming any particular
function or shape, and we obtain a tight constraint on the
foreground-orthogonal projection of the signal.
However, this leaves a wide range of possible full (non-orthogonal) signals
compatible with this constraint.

Our goal in this section is to differentiate between
these possible full signals and to find the most
likely underlying signal. To achieve this we parameterize
the signal with the FlexKnot method \citep{Vazquez2012,Millea2018,Heimersheim2022FRB} and apply Bayesian model selection to find
the most probable true signals. Our final result will be a
posterior distribution of signals, showing us which shapes 
are compatible with the data.

Even though multiple signals may fit the data equally well, we can
differentiate them based on their complexity. 
Simpler, more predictive models are more likely to be correct, whereas complex functions with many free parameters are susceptible to overfitting the data.
In Bayesian statistics, this effect is quantified as the Bayesian evidence
for a given model, and we can make use of this in the FlexKnot method.

\begin{figure}
	\centering
	\includegraphics[width=\linewidth]{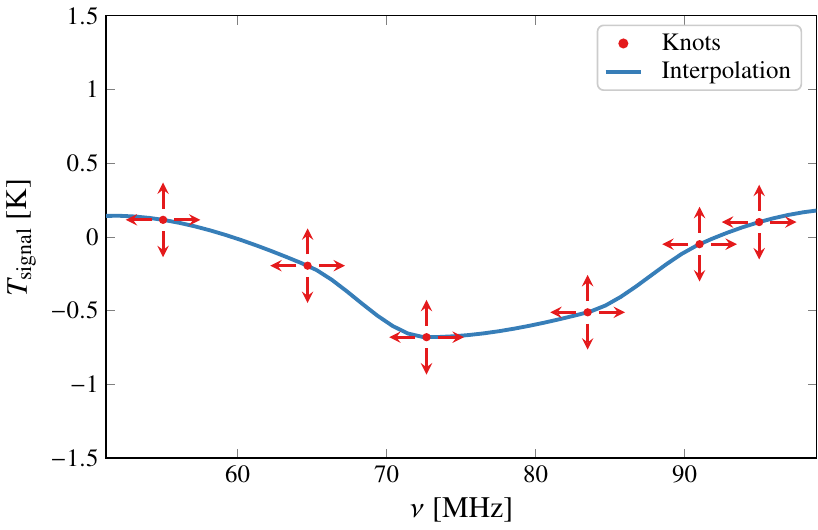}
	\includegraphics[width=\linewidth]{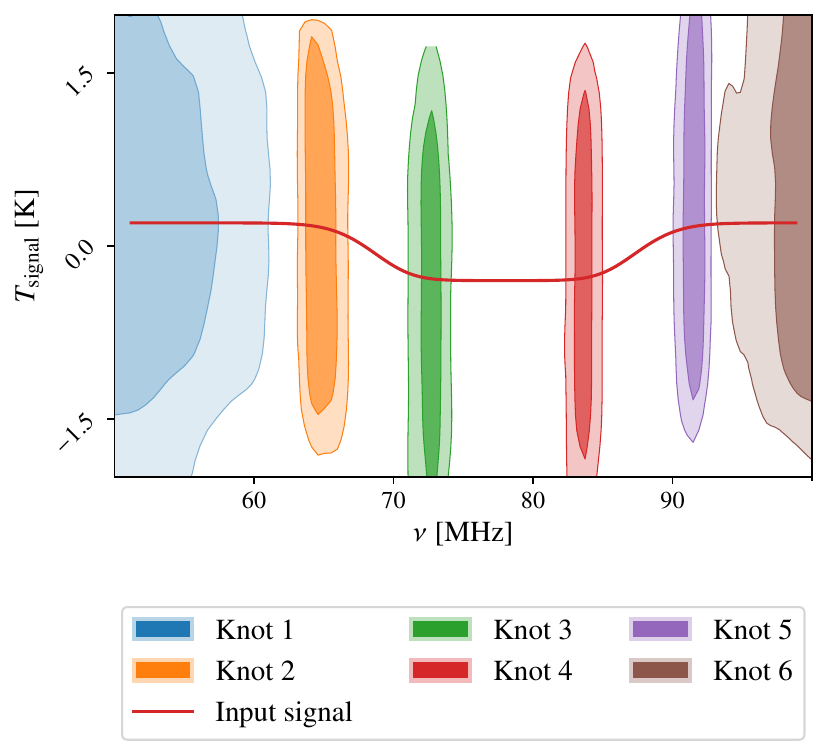}
	\caption[placeholder shortcaption]{Illustration of the FlexKnot parameterization $T_{\rm signal}(\nu)$.
	\textbf{Top:} An example FlexKnot with 6 interpolation points (knots, red)
	and the corresponding interpolated curve (blue). Only the knots (not the interpolation) are restricted by the priors to be within the range
	$\nu_i \in [50\,{\rm MHz}, 100\,{\rm MHz}]$ and $T_i \in [-2\,{\rm K}, 2\,{\rm K}]$.
    \textbf{Bottom:} An example posterior distribution of the knot coordinates
	$\left\{\nu_i, T_i\right\}$ for a FlexKnot run. This is the raw data
	which the signal posteriors (sections \ref{global:subsec:result_FK_mock} and \ref{global:subsec:result_FK_real}) are computed from.\footnotemark
	}
	\label{global:fig:illustration_knot_coordinates}
\end{figure}
\footnotetext{Note that the posterior distributions are very stretched
in the vertical direction, This is due to a degeneracy of the signal mean
with the foregrounds that we discuss in section
\ref{global:subsec:result_FK_mock}.}

We use the FlexKnot method to parameterize the signal
$T_\mathrm{signal}$ as a function of frequency $\nu$.
FlexKnot uses a family of flexible parameterizations
that are based on interpolation. The
first FlexKnot function is an interpolation between two
points, i.e. a straight line. The second FlexKnot function
adds a third point and interpolates between the three points,
and so on. We refer to these interpolation points as \textit{knots}.
Each function is parameterized by the coordinates $\left\{\nu_i, T_i\right\}$
of its knots, and the value at any other point $\nu$ is interpolated.
We illustrate this in the top panel of Figure \ref{global:fig:illustration_knot_coordinates}.

The interpolation function used between the knots is the piecewise cubic Hermite interpolation
polynomials (PCHIP)\footnote{We
use the scipy implementation \texttt{scipy.interpolate.pchip}
\citep{scipy2020}.}, following \citet{Millea2018}.
Alternative options are other spline variations
or even a linear interpolation. We have not found
meaningful differences between various interpolation
techniques, and chose the PCHIP interpolation to
avoid sharp the features that a linear interpolation
would introduce, making the functions easier
to interpret.
\rr{Note that, unlike \citet[][modelling the reionization history]{Heimersheim2022FRB}, we do not fix the first and last interpolation
knot to the minimum and maximum $y$-axis levels, as those minima
and maxima do not exist. Instead we extrapolate the cubic Hermite polynomials
before the first, and after the last interpolation knot.}

After setting up the FlexKnot parameterization we sample the
parameter space using Nested Sampling \citep{Skilling2004},
specifically we use the code \texttt{PolyChord}
\citep{Handley_2015,Handley_2015b}.
We perform multiple sampling runs for each number of knots
to ensure our results are not influenced by outlier runs,
averaging the posterior samples. In section \ref{global:subsec:result_FK_mock}
and \ref{global:subsec:result_FK_real} we show the evidences of individual
runs (specifically Figures \ref{global:fig:FK_mock_logZ} and \ref{global:fig:FK_real_logZ}) and show the scatter is small.
The parameter space
spans all knot coordinates and foreground parameters,
however as discussed further in section \ref{global:subsubsec:method_ortho_fg}
we only need to sample the knot coordinates. We choose
Nested Sampling so that we can compute the Bayesian
evidence $Z$ for each model, and we choose \texttt{PolyChord}
because it is well-suited for the high-dimensional parameter
space of FlexKnot models. Every Nested Sampling run gives us
the posterior distribution of knot positions for a given number
of knots; The lower panel of Figure
\ref{global:fig:illustration_knot_coordinates}
shows an example for $N_{\rm knots}=6$.
We use the package
\texttt{anesthetic} \citep{Handley2019b} to analyse the
Nested Sampling results. As the final visualization step 
do not show the distribution of knot coordinates, but the
distribution of functions by computing the interpolation for
every sample.
\rr{The reason for this is that the knot coordinates are not
directly interpretable, and the interpolation function is both,
what the data is ultimately sensitive to, and what we are
interested in.}
\rr{For this purpose we used} \texttt{fgivenx} \citep{Handley2018},
\rr{which yielded} the functional posterior distributions that we show in
section \ref{global:sec:results}.

We explore functions from $N_{\rm knots}=2$ to a maximum
of 15 knots. The cutoff was chosen empirically as the Bayesian
evidence for models declines after 6 knots, falling
to a negligible level of less than $10^{-6}$ of the
total evidence after 15 knots.
We would like to treat $N_{\rm knots}$ as a discrete parameter of
the model, but it is impractical to sample $N_{\rm knots}$ as a 
parameter. Instead, we run separate Nested Sampling runs for
each $N_{\rm knots}$ and average the results weighted by the
Bayesian evidence $Z$ of each run. 
This is equivalent to treating $N_{\rm knots}$ as a parameter and
marginalizing over its value. We can show this by writing the
posterior distribution of $T_{\rm signal}$ given
the data $\mathbf{d}$ and expanding the $N_{\rm knots}$ parameter. This
shows that $P(T_{\rm signal} \mid \mathbf{d})$ is equivalent to the
average over individual posteriors $P(T_{\rm signal} \mid N_{\rm knots}, \mathbf{d})$
weighted by the Bayesian evidence $Z(N_{\rm knots})$:
\begin{gather}
	P(T_{\rm signal} \mid \mathbf{d}) = \sum_{N_{\rm knots}} \underbrace{P(T_{\rm signal} \mid N_{\rm knots}, \mathbf{d})}_{\text{Individual posterior}} \underbrace{P(N_{\rm knots} \mid \mathbf{d})}_{\propto Z(N_{\rm knots})}.
\end{gather}
The same rewrite can be used for other quantities such as the foreground
parameter posterior.

\subsubsection{Orthogonalizing the foregrounds}
\label{global:subsubsec:method_ortho_fg}
There is one additional technique we want to introduce in this section.
While the FlexKnot method is \textit{not} limited to foreground-orthogonal
functions, we can make use of the orthogonalization to drastically speed-up
the sampling. We can analytically integrate out the foreground parameters
and reduce the parameter space we need to numerically sample by 5 dimensions.
We note that this idea is not limited to FlexKnot but can be applied to any
parameterization such as e.g. the flattened Gaussian or even physical models.
\\
In the following, we will show (i) how to find an orthogonal basis for the
foregrounds, and (ii) how to analytically solve the foreground parameter
marginalization.

\begin{figure}
	\centering
	\includegraphics[width=\linewidth]{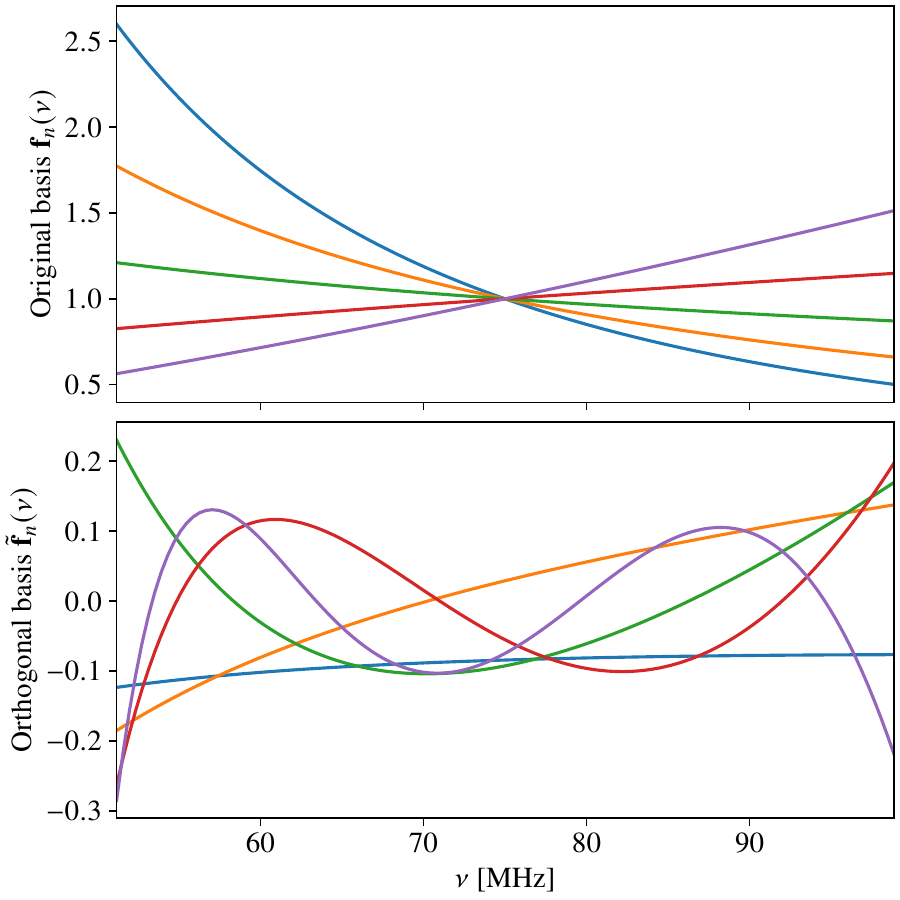}
	\caption{Original foreground basis functions $f_n(\nu)$ (top) and
	orthogonal foreground basis functions $\tilde f_n(\nu)$ (bottom).
	The new basis functions are orthogonal and normalized, and
	thus allow us to easily project signals into a foreground-orthogonal space.
	}
	\label{fig:illustration_foregrounds}
\end{figure}

\paragraph*{Orthonormal foreground basis:}
We want to express the foreground with new basis functions $\tilde f_n$
that are orthogonal and normalized
\begin{gather}
	T_{\rm fg}(\nu) = \sum_{n=0}^4 a_n \underbrace{\left(\frac{\nu}{\nu_c}\right)^{n-2.5}}_{f_n(\nu)}
	= \sum_{n=0}^4 \tilde a_n \tilde f_n(\nu),
	\label{global:eq:Tfg_orth}
\end{gather}
where the functions $\tilde f_n$ are orthonormal. We mark quantities
in the new basis with a tilde to distinguish them from the equivalent
quantities in the original basis.
The new foreground basis vectors are given by
$\mathbf{\tilde f}_n = \left(\tilde f_n(\nu_1), \dotsc, \tilde f_n(\nu_{123})\right)$.
Then $\mathbf{\tilde f}_n$ and $\mathbf{\tilde f}_m$ are orthonormal if and only if
$\mathbf{\tilde f}_n \boldsymbol\cdot \mathbf{\tilde f}_m = \delta_{nm}$.
To calculate this basis we define the $5\times 123$-dimensional
foreground matrix\footnote{This is the same matrix as the
design matrix from section \ref{global:subsubsec:method_GP_orthogonal}.}
$F_{ni} = f_n(\nu_i)$ and apply a singular value decomposition\footnote{Note
that the resulting matrices $\mathbf{U}$, $\mathbf{D}$
and $\mathbf{V}^T$ are $5\times5$, $5\times5$, and $5\times123$
dimensional, respectively; Figure \ref{fig:illustration_foregrounds}
shows the 5 123-dimensional rows of $\mathbf{F}$ (top) and $\mathbf{\tilde F} = \mathbf{V}^T$ (bottom).}
\begin{gather*}
	\mathbf{F} = \mathbf{U} \mathbf{D} \mathbf{V}^T
	\iff
	F_{ni} = \sum_{k=0}^4 U_{nk} D_{kk} (V^T)_{ki}
\end{gather*}
to obtain the desired orthogonal foreground basis $\tilde F_{ni} = (V^T)_{ni}$.
We can convert between the original foreground parameters $a_n$ and
the corresponding new basis parameters $\tilde a_n$ as
\begin{gather}
	a_n = \sum_{k=0}^4 \frac{U_{nk}}{D_{kk}} \tilde a_k
	\quad\mathrm{and}\quad
	\tilde a_n = D_{nn} \sum_{k=0}^4 U^T_{nk} a_k.
    \label{global:eq:basis_transformation_an}
\end{gather}
Figure \ref{fig:illustration_foregrounds} shows
a comparison of the original foreground basis functions $\mathbf{f}_n$ (top),
and the new orthogonal foreground basis functions
$\mathbf{\tilde f}_n$ (bottom). Note that orthogonal here refers to
the functions being orthogonal to each other, not orthogonal
to the foregrounds.

\paragraph*{Analytical foreground marginalization:} To fit a signal
model to the data one typically needs to simultaneously sample the
signal and foreground parameters. However, we found that for Gaussian
likelihoods and foreground models that are linear in their parameters,
such as our equation \eqref{global:eq:foreground_polynomial} or
equation (1) of \citet{Bowman2018}, there exists a shortcut. We
can analytically compute the posterior of the foreground parameters
conditional on the signal parameters. This means that we can sample
the foreground parameters analytically (negligible computational cost)
and need to sample only the signal parameters numerically.

We start by inserting the orthogonal foreground model, equation
\eqref{global:eq:Tfg_orth}, into the likelihood, equation
\eqref{global:eq:likelihood_simple}. This yields the
likelihood as a function of the FlexKnot parameters $\theta_{\rm FK}=
\{\nu_n, T_n\}$ and the foreground parameters $\theta_{\rm fg} = \{\tilde a_n\}$:
\begin{equation}
    P(\mathbf{d} \mid \theta_{\rm FK}, \theta_{\rm fg}) \propto \mathrm e^{\left(
        -\frac{
            \left(
                \boldsymbol{\xi} - \sum_n \tilde a_n \mathbf{\tilde f}_n
            \right)^2
        }{2\sigma_{\rm noise}^2}
        \right)}
\end{equation}
with the residuals vector
$\boldsymbol{\xi} = \mathbf{d} - \mathbf{T}_{\rm signal}(\theta_{\rm FK})$
(omitting the $\theta_{\rm FK}$ dependence for brevity).
We can simplify the exponent in two steps: First, we decompose $\boldsymbol{\xi}$ into
foreground-proportional terms with coefficients
$\mu_n = \boldsymbol{\xi} \boldsymbol\cdot \mathbf{\tilde f}_n$ and a foreground-orthogonal
remainder $\boldsymbol{\xi}_r$.
Secondly, we use the orthogonality of the bases to write the square of the sum
as a sum of squares, as all cross-terms vanish due to orthogonality.
\begin{align}
	\label{eq:likelihood_exponent_simplication}
	\left(\boldsymbol{\xi} - \sum_n \tilde a_n \mathbf{\tilde f_n}\right)^2
	=& \left( \sum_n (\mu_n- \tilde a_n)\, \mathbf{\tilde f}_n + \boldsymbol{\xi}_r\right)^2\\
    =& \sum_n \left((\mu_n- \tilde a_n)\, \mathbf{\tilde f}_n\right)^2
	+ \left(\boldsymbol{\xi}_r\right)^2
 \label{global:eq:factorization_likelihood_exponent}
\end{align}
Adding the foreground and FlexKnot prior terms gives us
\begin{align}
    P(\theta_{\rm FK}, \theta_{\rm fg} \mid \mathbf{d})
    \propto &\, \exp {\left(-\frac{(\mu_n-\tilde a_n)^2}{2\sigma_{\rm noise}^2}
    -\frac{\tilde a_n^2}{2 \sigma_{\tilde a_n}^2}\right)}
    \exp {\left(-\frac{\xi_r^2}{2\sigma^2_{\rm noise}}\right)}
    P(\theta_{\rm FK})
\end{align}
where the prior scale on foreground parameters $\sigma_{\tilde a_n}= 10^6\,\mathrm{K}$
is much larger than noise level $\sigma_{\rm noise}\ll 1\,\mathrm{K}$ and much larger
than the range allowed by the likelihood (on the order of $10^3\,\mathrm{K}$).
Thus, we can approximate the foreground prior term by a constant and obtain
\begin{gather}
    P(\theta_{\rm FK}, \theta_{\rm fg} \mid \mathbf{d})
    \propto \exp {\left(-\frac{(\mu_n-\tilde a_n)^2}{2\sigma^2_{\rm noise}}\right)}
    \exp {\left(-\frac{\xi_r^2}{2\sigma^2_{\rm noise}}\right)}
    P(\theta_{\rm FK}).
    \notag
\end{gather}
At this point we see the posterior has factorized into a foreground-dependent
Gaussian term, and a FlexKnot-dependent term. We can use this to both, analytically
sample the foreground parameters conditional on the FlexKnot parameters (simply a
Normal distribution with mean $\mu_n(\theta_{\rm FK})$ and scale $\sigma_{\rm noise}$), and to
marginalize over the foreground parameters as follows:
\begin{align}
    P(\theta_{\rm FK} \mid \mathbf{d}) &= \int_{-\infty}^{\infty}\mathrm{d}\theta_{\rm fg}\, P(\theta_{\rm FK}, \theta_{\rm fg} \mid \mathbf{d})
    \\
    &\propto \exp {\left(-\frac{\xi_r^2}{2\sigma^2_{\rm noise}}\right)}
    P(\theta_{\rm FK}).
\end{align}
We use this form in our implementation, sampling the FlexKnot parameters
$\theta_{\rm FK}$ numerically and sampling foreground parameters analytically
as derived parameters in \texttt{PolyChord}. We store both, the orthogonal
foreground parameters $\tilde a_n$ and the original foreground parameters $a_n$
in our chains, and will show the posterior distributions of both (Figure
\ref{global:fig:FK_mock_fgparams_tilde}, and \ref{global:fig:FK_mock_fgparams}
\& \ref{global:fig:FK_real_fgparams}, respectively).

We emphasize that, unlike for the GP analysis, this orthogonalization is not essential for the FlexKnot method to work. All it does is significantly improve the sampling efficiency, both by reducing the parameter space by 5 dimensions, and by letting us skip sampling over the degenerate foreground parameters.

\rr{
\subsubsection{Mean subtraction}
\label{global:subsubsec:method_FK_meansubtraction}
We apply one important post-processing step to the final FlexKnot
signals, which is to normalize all functions to have zero\footnote{Zero
is an arbitrary choice for illustration purposes.} mean. This
is necessary because, while FlexKnot can eliminate almost the entire
foreground degeneracy that affects the shape of the signal, it cannot
eliminate the degeneracy between the absolute signal level and the foreground level.

This is because (i) the foregrounds can combine to create an almost constant offset in the signal, as demonstrated by the blue line in the lower panel of Figure \ref{fig:illustration_foregrounds}.
And because (ii) a FlexKnot parameterization can also allow for a constant offset by shifting all knot coordinates vertically.
The latter does not incur a complexity penalty, thus creating a degeneracy between the signal level and the foreground level that cannot be removed by the FlexKnot method.

The post-processing step allows us to visualize the good constraint on the shape and contrast of the signal.
In terms of scientific results, this means we cannot constrain quantities that depend on the absolute signal level but it leaves inferences about the shape of the signal unaffected.
Constraints on absolute signal levels can be made possible by adding more assumptions, such as fixing the value of the signal at a given frequency based on a physical model, but this is beyond the scope of this work.

Note that a similar technique does not solve the degeneracy of the Gaussian Process fit
seen in Figure \ref{global:fig:GP_nonorthog}. This is because the GP already penalizes
constant offsets of the fit, and an additional mean subtraction would be less useful.
We explicitly show this in appendix \ref{global:subsec:appendix_GPmeanzero}.
}

\subsubsection{FlexKnot priors}
\label{global:subsubsec:method_FK_priors}
We choose simple priors on the FlexKnot
parameters, i.e. the coordinates of the interpolation points. We allow the
frequency coordinates $\nu_i$ to lie between 50\,MHz and 100\,MHz (the
frequency range corresponding to the EDGES low-band observations),
and the temperature coordinates $T_i$ to lie between
$-2$\,K and $2$\,K (the largest signal amplitude we consider\footnote{Because we know 
there exist signals $<1$\,K that can fit the data
\citep[as demonstrated in][]{Bowman2018} we do not consider much larger signals.}).
The temperature prior is uniform (flat), for the frequency prior we use a forced
identifyability prior \citep{Handley2015b}. This means we sample the frequency
coordinates as though they were drawn from a uniform (flat) prior, and
then sorted so that the knot coordinates are always in ascending order.
This gives the same result as a uniform (flat) prior on $\nu_i$, but
avoids the combinatorial explosion of having to sample every possible
permutation of the knot coordinates.
As for the number of interpolation knots $N_{\rm knots}$, we use a discrete
uniform prior from the minimum of 2 knots up to 15 knots; we notice
empirically that the evidence for models with more than 15 knots
is negligible.
As prior on the foreground parameters we adopt a Normal distribution
$\mathcal{N}\left(0\,\mathrm{K}, 10^6\,\mathrm{K}\right)$ over the transformed foreground
parameters $\tilde a_n$ (see section \ref{global:subsubsec:method_ortho_fg}).
Given that the actual values lie in the $\pm10^3\,\mathrm{K}$ range\footnote{We will
see this in section \ref{global:sec:results}, specifically in Figures
\ref{global:fig:FK_mock_fgparams} and \ref{global:fig:FK_real_fgparams}.},
this is effectively an uninformative and approximately flat prior.
The difference in priors between this choice and a prior on $a_n$ (as chosen
for the Gaussian Process) is negligible, as the priors have such a small effect on our results.
The noise level $\sigma_{\rm noise}$ is a free parameter as well, here we choose
the same prior as in the Gaussian Process run,
a non-negative Normal distribution
$\mathcal{N}^+(0\,\mathrm{K}, 1\,\mathrm{K})$.

In addition to these,
we also
consider additional priors to encode our prior knowledge about the
signal. In our analysis we consider either: a uniform prior
$[-2\,\mathrm K, 2\,\mathrm K]$ on the signal values, which affects the posterior
only slightly beyond the existing knot coordinate priors, a Gaussian prior 
$0\,\mathrm{K}\pm1\,\mathrm K$ on the signal \textit{contrast} (the difference between the
highest and lowest signal value), or a Gaussian prior on every signal point.
We use the second option of a Gaussian prior on the
signal contrast for the rest of our analysis and show the effects of
other choices in appendix \ref{global:subsec:appendix_FKpriors}.

\section{Results}
\label{global:sec:results}

We apply the Gaussian Process and FlexKnot methods to both, the
synthetic data, where we know the ground truth, and the 
EDGES low-band data from \citet{Bowman2018} to test our methods on a real measurement. \rr{We will first discuss the GP fit (section \ref{global:subsec:result_GP}) to both data sets, then the FlexKnot fit
to the synthetic data (section \ref{global:subsec:result_FK_mock})
and EDGES low-band data (section \ref{global:subsec:result_FK_real}).}

\subsection{{Gaussian Process}}
\label{global:subsec:result_GP}
\begin{figure}
	\centering
	\includegraphics[width=\linewidth]{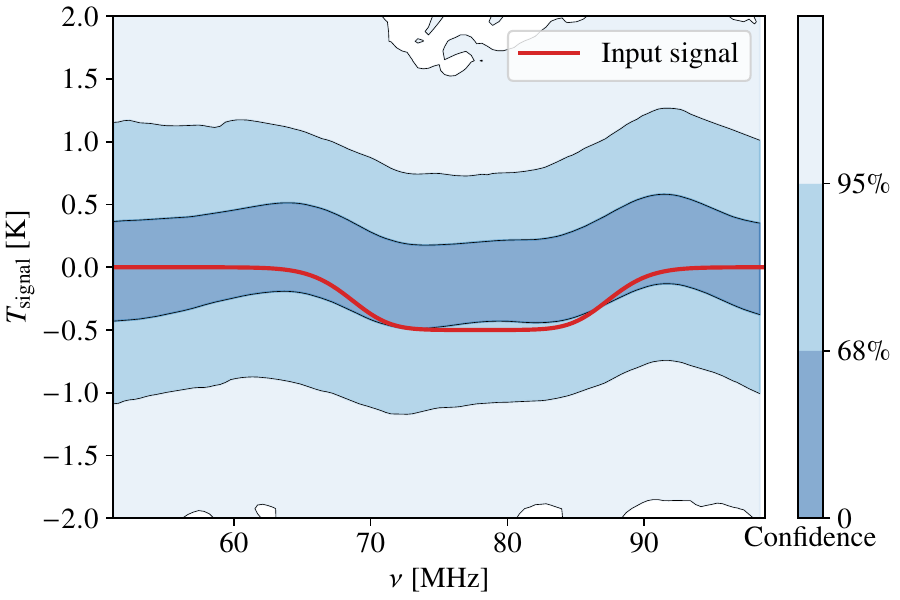}
	\includegraphics[width=\linewidth]{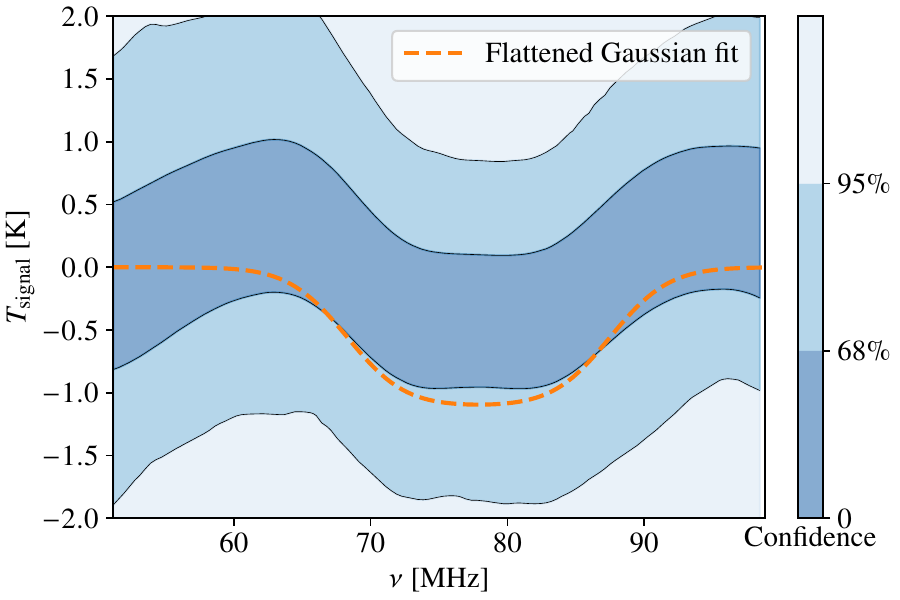}
	\caption{\rr{Gaussian Process posterior contours without orthogonalization.
	The blue contours show the posterior distribution of signals for the
	synthetic dataset (top) and the EDGES low-band data (bottom). The red line
	(top) indicates the input signal (ground truth) for the synthetic data,
	the orange dashed line (bottom) shows the flattened Gaussian fit to the
	EDGES data. The Gaussian Process contours for the synthetic data are
	consistent with the input, but both Gaussian Process posteriors are
	rather wide and unconstraining due to their degeneracy with the foreground
	term.}}
	\label{global:fig:GP_nonorthog}
\end{figure}

\rr{As discussed in section \ref{global:subsec:method_GP}, we fit both, an
 unconstrained GP and a foreground-orthogonal GP. Extracting the signal with an
unconstrained fit proves difficult, but we can extract the foreground-orthogonal part
of the signal with a high degree of confidence. We will start this section with
that result, and show the corresponding results for the unconstrained GP in
Figure \ref{global:fig:GP_nonorthog}.}

\begin{figure}
	\centering
	\includegraphics[width=\linewidth]{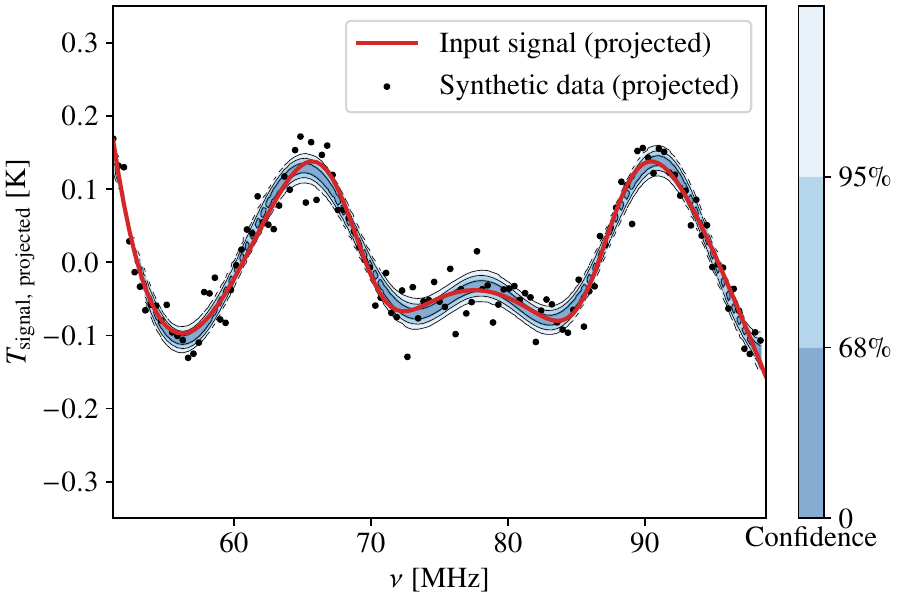}
	\caption{Gaussian Process fit to the synthetic data set
	introduced in section \ref{global:subsec:method_data}.
	The blue contours indicate the 68\%, 95\% and 99.7\%
	credible intervals of the temperature values
	$T_\mathrm{signal,\ projected}$ at every frequency,
	derived from the Gaussian Process posterior. The red line shows
	the foreground-orthogonal component of the true signal (see
	section \ref{global:subsec:method_data} for details) and the black
	dots show the foreground-orthogonal component of the synthetic data vector.
	The figure shows that the Gaussian Process correctly recovers
	the input signal shape
	and its uncertainty is consistent with the scatter in the data.
	}
	\label{global:fig:GP_mockresult}
\end{figure}

We show the \rr{foreground-orthogonal} Gaussian Process posterior fit
to the synthetic data
in Figure \ref{global:fig:GP_mockresult}. The plot shows the
posterior distribution (blue contours) of the Gaussian
Process that is (by construction)
orthogonal to the foreground component. The shades of blue show the
$1\sigma$, $2\sigma$ and $3\sigma$ contours, corresponding to 68\%, 95\% and 99.7\%
credibility intervals.
These constraints on the foreground-orthogonal components show us
which data features need to be explained by
systematics or an astrophysical model.
For comparison, we also show the projected ground truth signal (red line) and
a projection of the synthetic data (black points) that this fit is based on.
These projections are the foreground-orthogonal components of the ground
truth signal and data, respectively, using the orthogonalization discussed
in section \ref{global:subsubsec:method_ortho_fg}.
We see that the GP posterior matches
the ground truth signal very well, and its width is consistent with
the scatter in the data points.
The posterior for the noise level (a free parameter) is
$\sigma_\textrm{noise}=0.027\,\mathrm{K}\pm0.002\,\mathrm{K}$, which is 
consistent with the input noise level of $0.025$\,K.\footnote{The noise level of the foreground-orthogonal component is expected to be
the same as the full signal, as the foreground-parallel component cannot absorb
random thermal noise.}
As for the other GP parameters, we give mean values and 95\% credible
intervals to account for the asymmetry of the distribution.
The amplitude $\sigma_f$ has mean $0.93$\,K and lies within
$[0.2\,\mathrm{K}, 2.1\,\mathrm{K}]$; the kernel length scale
$\ell$ has mean 7.3\,MHz and lies within $\in [5.5\,\mathrm{MHz}, 8.7\,\mathrm{MHz}]$.

\begin{figure}
	\centering
	\includegraphics[width=\linewidth]{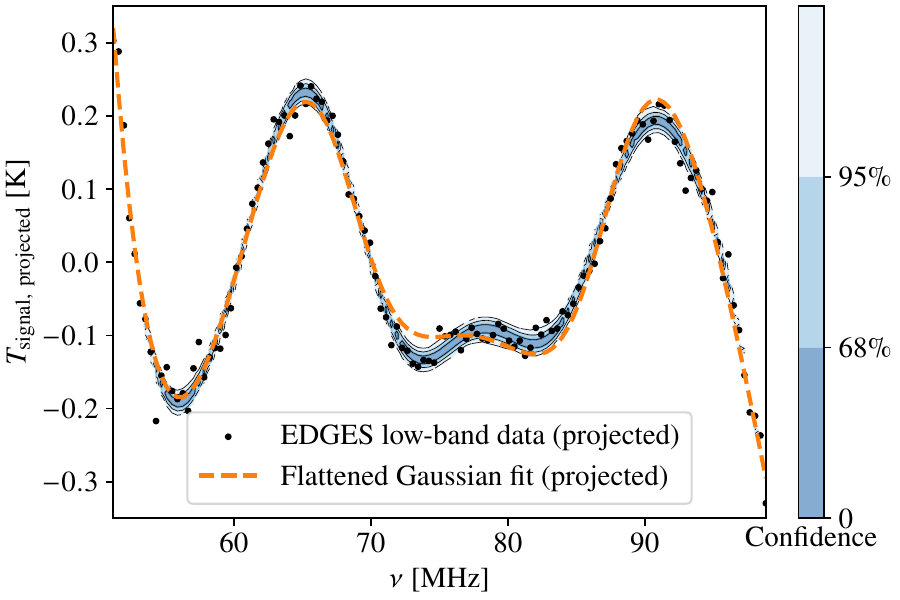}
	\caption{Gaussian Process fit to the EDGES low-band
	data \citep[from][]{Bowman2018}.
	We show posterior credible intervals as blue contours,
	and the projected data points in black. We do not have a ground truth
	as the real cosmological or systematic signal present is unknown.
	However, we can show a flattened Gaussian fit (orange dashed line,
	equation \ref{global:eq:flattened_gaussian}) following \citet{Bowman2018}
	for comparison.
	This shows us which features in the signal the flattened
	Gaussian cannot explain, and we see multiple points where the GP and
	flattened Gaussian fits deviate by more than 3$\sigma$.
	\\
	Note that the orange dashed line here is a \textit{fit} (varying $A$, $\nu_0$,
	$w$, and $\tau$) of equation
	\eqref{global:eq:flattened_gaussian} to the EDGES data,
	while the red solid line in Figure \ref{global:fig:GP_mockresult} is
	the known input signal for the synthetic data. Both look alike because the
	input signal for the synthetic data was chosen to be similar to the
	best-fitting signal in \citet{Bowman2018}.
	Also note that the shape of the flattened Gaussian looks unfamiliar
	because we are only showing its foreground-orthogonal component.
	}
	\label{global:fig:GP_realresult}
\end{figure}

Applying the same method to the EDGES low-band data from \citet{Bowman2018}
we obtain the posterior shown in Figure \ref{global:fig:GP_realresult}.
Again we show the posterior contours in blue, and the projected data points
in black, but in this case we do not know the true signal.
An interesting comparison is to see how well a flattened Gaussian signal
\citep[as reported in][]{Bowman2018}
can explain the data, or equivalently, how well it can explain
the foreground-orthogonal component shown here. We show a (projected) fit
of the flattened Gaussian model (equation \ref{global:eq:flattened_gaussian})
as an orange dashed line, and see that although the flattened Gaussian fit
largely overlaps with the GP posterior, it significantly deviates from the GP contours around 74\,MHz, 82\,MHz, and 91\,MHz.

The Gaussian Process fit posterior for the noise level is
$\sigma_\mathrm{noise} = 0.021\,\mathrm{K}\pm 0.002\,\mathrm{K}$, somewhat lower than found
in \citet{Bowman2018}, which is expected, as the Gaussian Process
fits the data more accurately. For the other parameters, we again
give the mean value and 95\% credible intervals:
The amplitude mean is $1.95$\,K with the credible interval
$\sigma_f \in [0.4\,\mathrm{K}, 3.8\,\mathrm{K}]$;
the mean kernel length scale is 7.23\,MHz with interval
$\ell \in [5.6\,\mathrm{MHz}, 8.9\,\mathrm{MHz}]$.

We want to comment on the sharp features at 70\,MHz and 90\,MHz in
Figures \ref{global:fig:GP_mockresult} and \ref{global:fig:GP_realresult}.
These are reminiscent of the features found in the original EDGES analysis
\citep{Bowman2018} and demonstrate an example of a feature that cannot
be absorbed by the foregrounds, and thus is present in both analyses.\footnote{The
same feature appears in the synthetic data analysis because that data is generated
with a similar profile as the EDGES flattened Gaussian fit, as discussed in section \ref{global:subsec:method_data}.}
Note however that this feature does not necessarily have to be present as a
sharp rise and fall like it is here: In section \ref{global:subsec:result_FK_real}
(specifically Figure \ref{global:fig:FK_real_lines}) we will see signal
shapes that contain the same foreground-orthogonal component but look
visually different. This highlights why the foreground-orthogonal projection
is such a useful tool for comparing whether different signals are consistent
with the data, or with each other.

\rr{As a point of comparison we also show the fit
of an unconstrained Gaussian Process to both data sets in
Figure \ref{global:fig:GP_nonorthog}. The posterior constraints show large uncertainties
(blue contour bands) which are caused by the degeneracy with foregrounds. These
constraints depend somewhat on the choice of priors, but the qualitative result
remains the same.}

\subsection{FlexKnot signal fit to synthetic data}
\label{global:subsec:result_FK_mock}
Unlike our Gaussian Process, the FlexKnot method \textit{can} reliably produce 
a posterior for the full signal. As discussed in section
\ref{global:subsec:method_FK}, the reason this is possible is that
the FlexKnot method has an implicit prior towards simpler, more predictive
functions due to the Bayesian evidence weighting. This is the distinguishing
criterion between multiple signals that fit the data equally well: while their
foreground-orthogonal projection can be the same, the full signals
differ in complexity. The FlexKnot method will find the simplest signals
among the infinite number of signals that are consistent with the data.
While the simplest signal is not necessarily the true signal, we expect
it to be at least instructive, and notice that the method does in fact
recover the ground truth for the synthetic data.

\begin{figure}
	\centering
	\includegraphics[width=\linewidth]{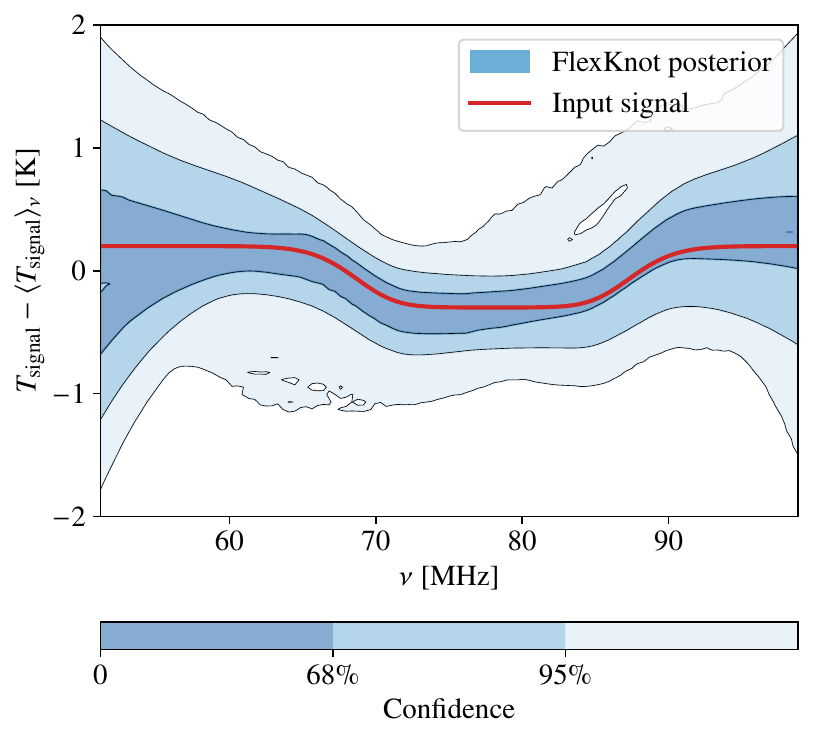}
	\includegraphics[width=\linewidth]{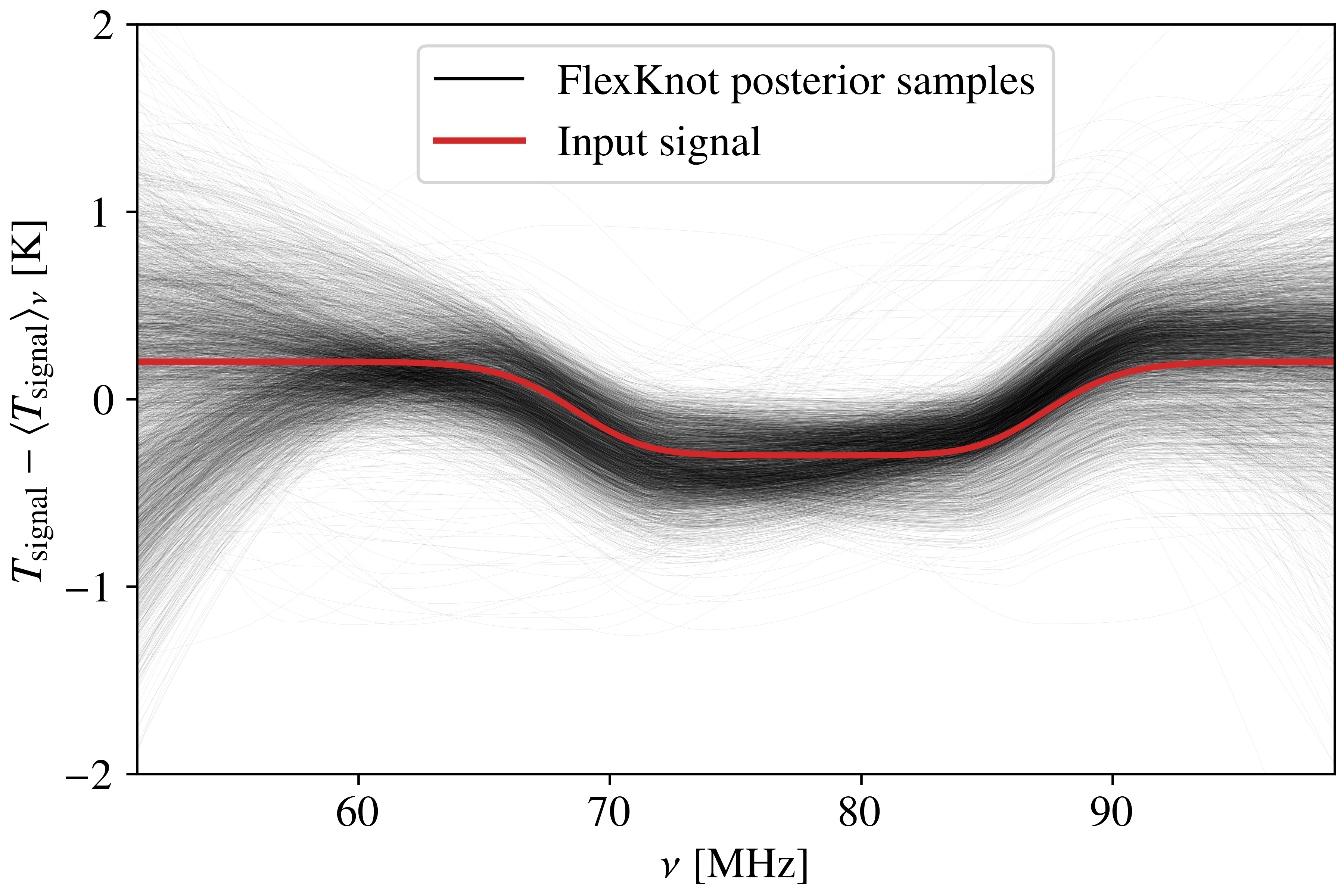}
	\caption{FlexKnot fit to our synthetic data set, visualized as posterior
	contours (top) and random posterior samples (bottom).
	These plots use the full signal $T_{\rm signal}$, rather than the
	foreground-orthogonal projection we showed for the GP fit. The only
	post-processing we apply is to shift the signal to have zero mean
	$T_{\rm signal} - \langle T_{\rm signal} \rangle_\nu$.
	Both plots show the input signal (red line) as a ground truth that
	the synthetic data was generated from.
	\textbf{Top:} Credible intervals of the temperature value at each
	frequency $\nu$, visualized as blue contours correspond to 68\%, 95\%
	and 99.7\% credible intervals respectively. We see that the ground
	truth (red line, synthetic data input signal) is recovered within
	the 68\% credible region of the posterior.
	\textbf{Bottom:} Random samples drawn proportional to the FlexKnot posterior.
	Every sample is a function $T(\nu)$  shown as a thin black line. This 
	visualization gives us a better view of the functional shapes contained
	in the posterior. We see that many samples have a similar shape to the
	ground truth (red line), although a fraction of posterior samples spread
	out at low frequencies and do not recover the flat nature of the ground truth.
	Overall though we could infer the underlying signal from observing the
	posterior samples.}
	\label{global:fig:FK_mock}
\end{figure}

First, we plot the posterior contours of the FlexKnot functional
posterior, showing the posterior of the signal value at
every frequency. The top panel of Figure \ref{global:fig:FK_mock}
shows credible intervals (blue contours) along with the ground truth signal
(red line) used to generate the synthetic data. We see
that the contours are compatible with the input signal 
and indeed suggest the correct shape. Overall we see a much larger
uncertainty than in the orthogonal Gaussian Process fit, which is expected
since here we are fitting the full signal, not just the foreground-orthogonal
component. The main source of uncertainty is the degeneracy with the
foregrounds.

We make one post-processing adjustment to the posteriors for
all FlexKnot plots in this section: we shift all signals to have zero mean. This
is because one of the foreground components is
an approximately constant signal,
and without additional priors, we cannot resolve that degeneracy
since the constant does not add any complexity.
However, we are mostly interested in the shape of the signal here,
and not its absolute level, thus this is not an issue in our analysis.
Note that this is the same effect that causes the FlexKnot knot coordinate
posteriors in Figure \ref{global:fig:illustration_knot_coordinates} to
be stretched out vertically. In Figure
\ref{global:fig:appendix_meanzero_demo_DiffGaussian} we show
the posterior functions without this
mean-zero shift. In the same figure, we also show the
foreground-orthogonal projection of the FlexKnot
posterior for comparison with the GP results.

In the posterior contour plot (top panel
of Figure \ref{global:fig:FK_mock}),
it can be hard to determine the shapes of
individual functions, so in addition, we want to
look at a representative sample (drawn
proportional to the posterior) of signals.
The bottom panel of Figure \ref{global:fig:FK_mock}
shows $\approx 4,000$ of such function samples, and allows us to get an intuition
for the functional shapes. Here we can see that almost all functions
have a flattened absorption trough, similar to the ground truth signal
(red line). On the low-frequency tail, we see the posterior spreading out,
as well as at the high-frequency tail (although to a lesser degree).

\begin{figure}
	\centering
	\includegraphics[width=\linewidth]{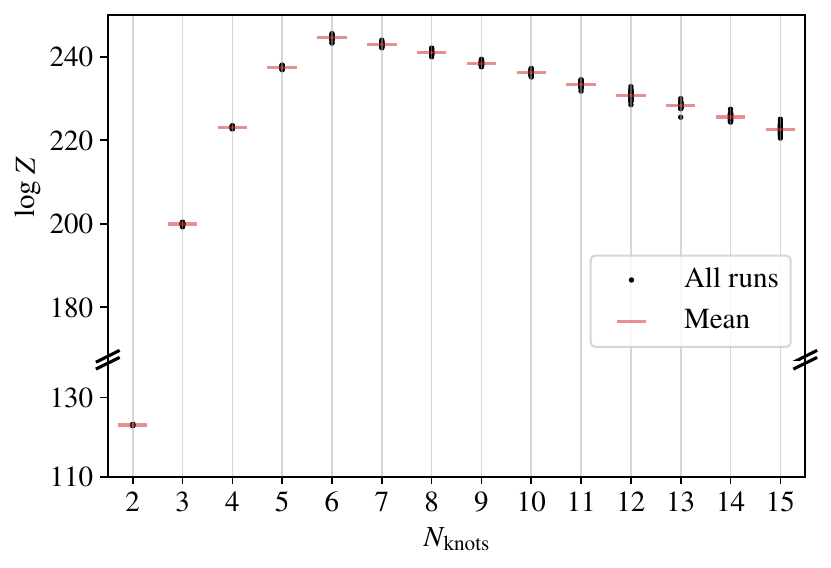}
	\caption{Bayesian evidence (logarithmic, base $e$) as a
	function of the number of knots for the FlexKnot fit to the synthetic data set. Marginalizing over the number of
	knots corresponds to weighing runs by their evidence $Z$ so this plot
	shows us which runs account for the largest amount of posterior
	mass. For each $N_{\rm knots}$ value we perform multiple Nested Sampling
	runs (shown as black dots) to account for outliers, and take the mean (red markers) of the $\log Z$ values
	since $Z$ is log-normal distributed \citep{Handley2015b}.
	}
	\label{global:fig:FK_mock_logZ}
\end{figure}

The posterior distribution is marginalized over the number of knots.
As discussed in section \ref{global:subsec:method_FK}, this is an average
over all runs, weighted by the Bayesian evidence of each
run.\footnote{The
weights are normalized to sum to 1 while preserving the
relative weight between runs.}
We show the evidence as a function of $N_\mathrm{knots}$ in Figure
\ref{global:fig:FK_mock_logZ}. 
We see that the evidence rises sharply until we reach the
peak at $N_{\rm knots} = 6$, and falls off continuously
after that. 
Both effects are what we would expect: the sharp rise since low-$N_{\rm knots}$ models cannot adequately fit the data, and the fall off since high-$N_{\rm knots}$ models are more complex than necessary and less predictive, thus
being weakly penalized by the Bayesian evidence.
At the largest number we explore, $N_{\rm knots} = 15$, the evidence has
fallen to $Z_{15}\approx \mathrm e^{223}$ which is smaller than the peak evidence ($Z_6\approx \mathrm e^{245}$)
by a factor of $>10^9$.
Thus, we expect larger numbers of knots to be negligible, and truncate
our analysis after $N_{\rm knots} = 15$.

\begin{figure*}
	\centering
	\includegraphics[width=0.8\linewidth]{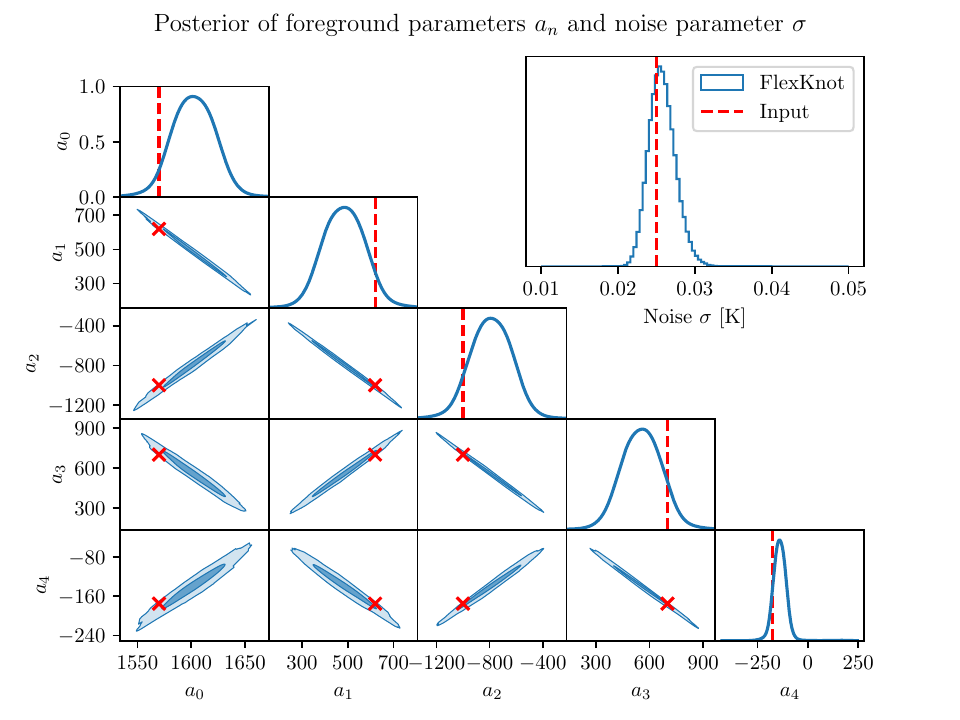}
	\caption{The posterior distribution of noise and foreground parameters for
	the FlexKnot signal fit to the synthetic data set. We find the expected
	degenerate distribution of foreground parameters but recover the input
	values (this is more visible in the upper panel of Figure 
    \ref{global:fig:FK_mock_fgparams_tilde} in the Appendix).
    Note that the degeneracy is not a problem for our Nested Sampling
	due to the foreground orthogonalization discussed in
	section \ref{global:subsubsec:method_ortho_fg}.
	The noise posterior distribution shown in the top right insert
	is consistent with the input noise level (red dashed line).
	}
	\label{global:fig:FK_mock_fgparams}
\end{figure*}

Finally, the noise level and posteriors of the foreground parameters are shown in Figure \ref{global:fig:FK_mock_fgparams}.
As expected, the foreground parameters are highly correlated.
We note that the transformed foreground parameters $\tilde a_n$ (see section \ref{global:subsubsec:method_ortho_fg}) are significantly less correlated, as demonstrated in Figure \ref{global:fig:FK_mock_fgparams_tilde}.
In both Figures, we see that the foreground posterior parameters are consistent with the input parameters at the 95\% confidence level.
We find a noise level of $\sigma_{\rm noise} = 0.026\,\mathrm{K} \pm 0.001\,$K, also consistent with $0.025\,$K noise added to the synthetic data.

As discussed in section \ref{global:subsec:method_FK} we have chosen
a Gaussian prior on the largest difference in the signal values
for our FlexKnot implementation. Different choices of priors can
affect the posterior if the signals differ significantly, but the
effect on the synthetic data set is small. We show posterior plots
for other prior choices in appendix \ref{global:subsec:appendix_FKpriors},
for both the synthetic data and the EDGES low-band data.

\subsection{FlexKnot signal fit to EDGES low-band data}
\label{global:subsec:result_FK_real}

\begin{figure}
	\centering
		\includegraphics[width=\linewidth]{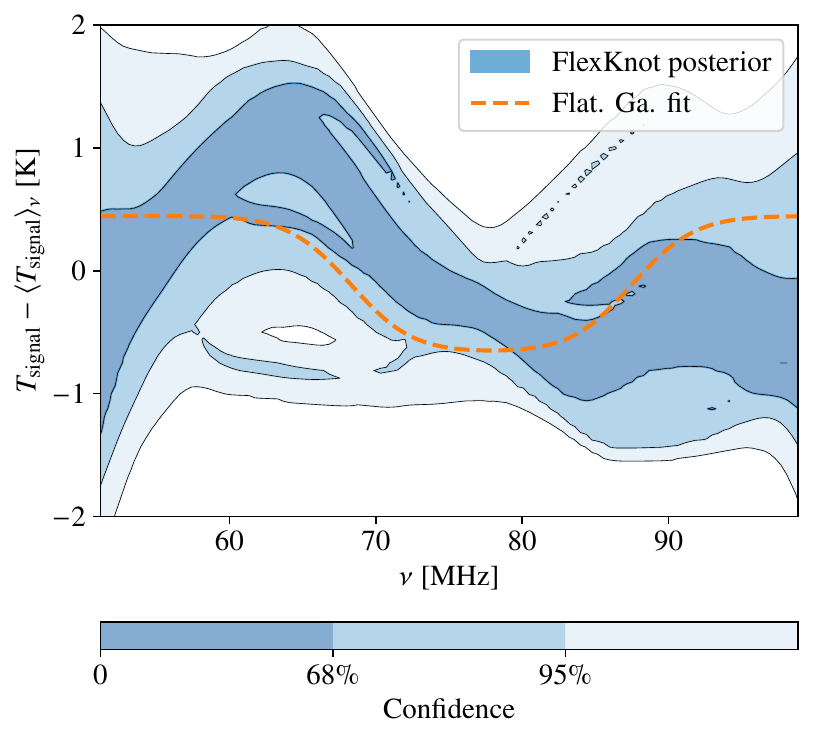}
		\includegraphics[width=\linewidth]{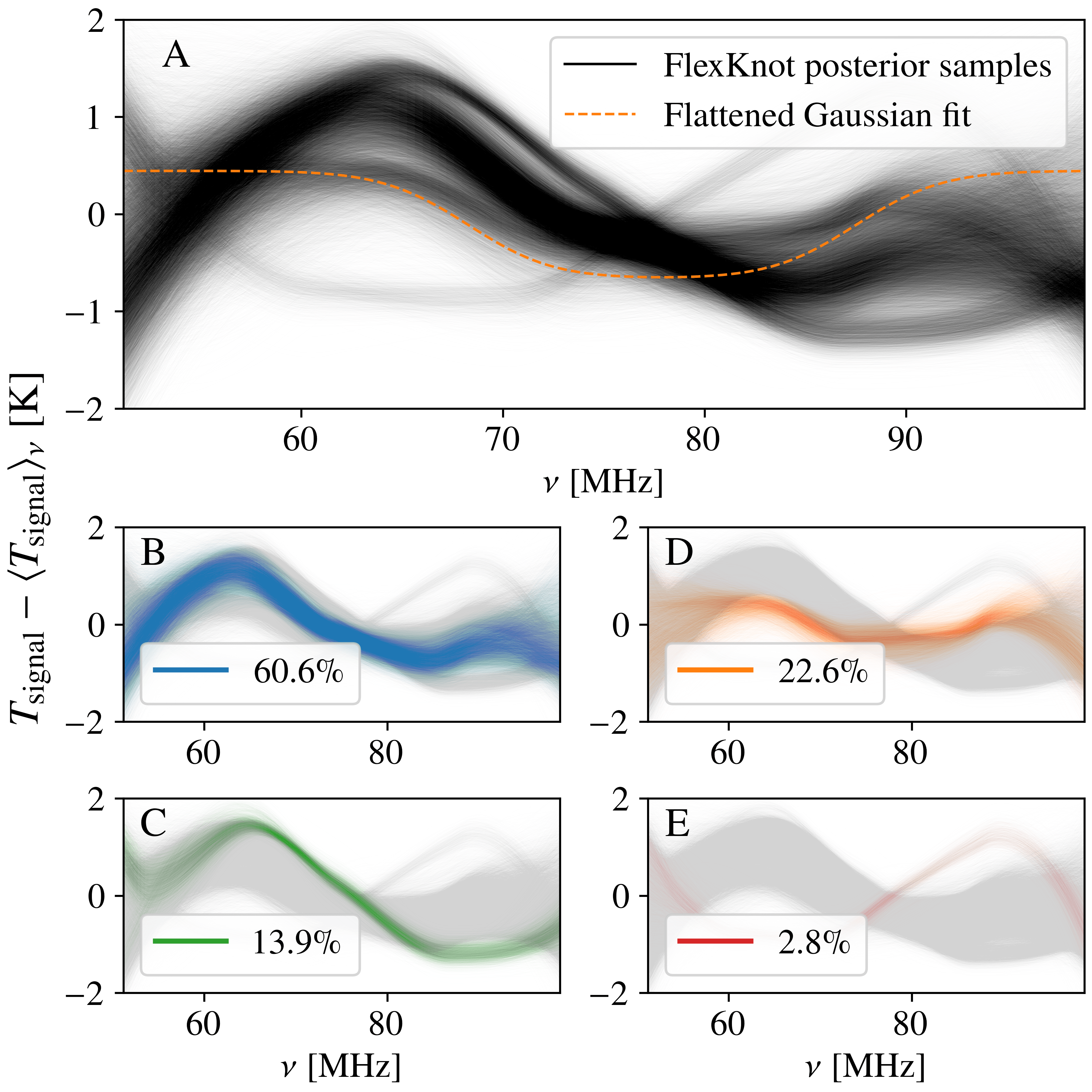}
	\caption{FlexKnot posteriors for the EDGES low-band data set
	\citep[][]{Bowman2018}, visualized as credible intervals (top) and
	posterior samples (bottom). The plots show the full signal except
	for a mean-zero shift, as described in section \ref{global:subsec:result_FK_mock}.
	\textbf{Top:} Credible intervals of the temperature
	$T_{\rm signal} - \langle T_{\rm signal} \rangle_\nu$
	at every frequency (blue contours) derived from the FlexKnot fit,
	along with the best-fitting flattened Gaussian model (orange dashed line,
	equation \ref{global:eq:flattened_gaussian}) for comparison. While we
	can recognize the main posterior mode from this plot, the contours
	indicate a multi-modal distribution and don't describe the
	posterior in a useful way.
	\textbf{Bottom:} Random samples drawn from the FlexKnot posterior. We
	show five subplots in this panel: Subpanel \textbf{A} shows samples
	drawn from the whole posterior (black), as well as the 
	best-fitting flattened Gaussian model (orange dashed line) again for comparison.
	The posterior samples (black line) clearly show the multi-modal nature
	of the posterior. We can distinguish four modes in the posterior by eye, and
	colour these in the subpanels below. Subpanel \textbf{B} highlights
	the largest mode (blue, 61\% of the posterior), followed by
	subpanel \textbf{C} (green, 14\%), subpanel
	\textbf{D} (orange, 23\%), and subpanel \textbf{E} (red, 2.8\%). In each
	panel, we show all posterior samples in grey and highlight the respective
	mode in colour.
	\label{global:fig:FK_real_contours}
	\label{global:fig:FK_real_lines}
	}
\end{figure}

Now we apply the FlexKnot analysis to the EDGES low-band data from \citet{Bowman2018}.
In Figure \ref{global:fig:FK_real_contours} (top panel) we show the
resulting posterior contours along with the flattened Gaussian fit for comparison.
The posterior contours are more complex compared to the synthetic data (which
only contain the flattened Gaussian signal, a foreground, and random noise) and appear
multi-modal. The credible intervals are rather large, as the different
modes cover different parts of the temperature range.

To better visualize this posterior distribution we again turn to a plot
of randomly drawn posterior function samples, shown in the bottom
panels of Figure \ref{global:fig:FK_real_lines}.
We can see a clear distinction and
observe the individual signal modes in this plot.\footnote{We
have reproduced this plot with multiple Nested Sampling chains to confirm
that the modes correspond to the posterior shape and are not an
artefact of the sampling.}
We show the same posterior samples in five subpanels
of Figure \ref{global:fig:FK_real_lines}: subpanel \textbf{A} shows
all samples in black, and again the best-fitting flattened Gaussian for comparison. The next four subpanels (\textbf{B} through \textbf{E}) each
highlight a subset of the posterior samples in colour to illustrate
the different apparent modes in the posterior.
This ad-hoc classification is based on the signal slope between 73 and 84\,MHz
and should not be treated as a rigorous classification, and thus the numbers we give below are only to be taken as estimates.

We notice that the majority of the posterior samples (subpanel \textbf{B},
$\sim 61\%$) follow an oscillatory pattern with a peak around 65\,MHz,
a trough around 85\,MHz, and another peak around 95\,MHz.
\\
A smaller fraction
of the data (subpanel \textbf{C}, $\sim 14\%$) shows a similar shape with an
additional trough around 55\,MHz and without the 95\,MHz peak.
Neither of these two signal modes is compatible with the
flattened Gaussian shape, and neither would be detected in
an analysis that forcefully assumes such a shape.
\\
The third mode (subpanel \textbf{D}, $\sim 23\%$) resembles the
signal found by \citet{Bowman2018} with a trough at 78\,MHz,
except for the tails at low and high frequencies that are not flat
in our case. We also note that the flattened Gaussian best-fitting
profile (orange dashed line in subpanel \textbf{A}) shows a much deeper trough
that is not recovered by the FlexKnot fits.
\\
Finally, the small fourth family of posterior samples (subpanel \textbf{E},
$\sim 2.8\%$) follows an inverted oscillatory pattern, with a trough
around 65\,MHz and a peak around 90\,MHz. However, it is much less probable
than the other modes, and only highlighted here for its distinct shape.

The balance between the different signal modes is directly influenced by
any priors chosen for the analysis. As we show in Appendix
\ref{global:subsec:appendix_FKpriors}, the prior
choice does not particularly affect the shapes, but does change the relative
probabilities of the different modes. In particular, the third mode (subpanel \textbf{D}) has a lower contrast (difference between maximum and minimum
of the signal) and thus is preferred by the Gaussian prior (see Figure
\ref{global:fig:appendix_FKpriors}).

\begin{figure}
	\centering
	\includegraphics[width=\linewidth]{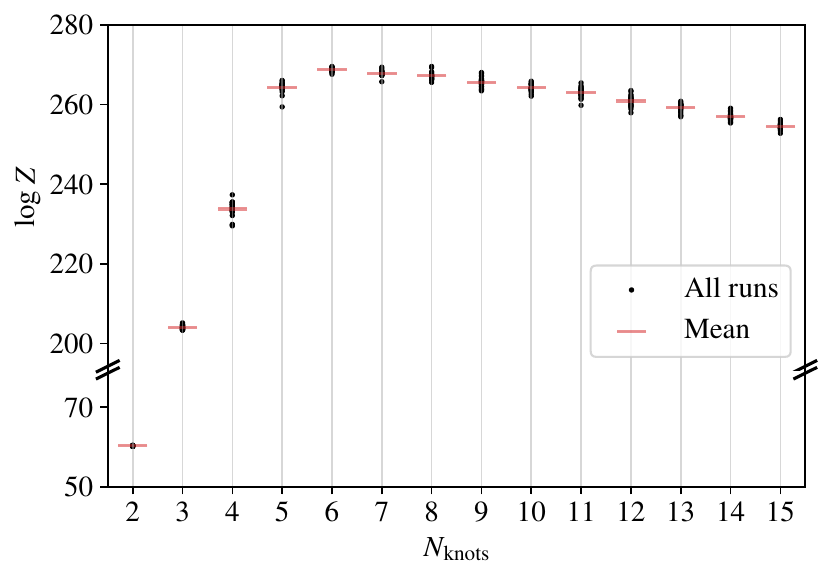}
	\caption{Log evidence as a function of $N_{\rm knots}$ for FlexKnot results on
    the EDGES low-band data \citep[from][]{Bowman2018} (note the split $y$-axis). Again we show individual runs (black dots) as well as the mean
	(red markers) for every $N_{\rm knots}$ value. The evidence again peaks at $N_{\rm knots} = 6$
	but falls off less strongly than for the synthetic data
	(Figure \ref{global:fig:FK_mock_logZ}). The evidence
	still falls off by more than a factor of $10^6$ making
	$N_{\rm knots} > 15$ negligible. The slower fall-off suggests that the
	signal contained in the EDGES data is more complex than the flattened
	Gaussian in our synthetic data.}
	\label{global:fig:FK_real_logZ}
\end{figure}

We also show the Bayesian evidence as a function of $N_{\rm knots}$
in Figure \ref{global:fig:FK_real_logZ}. The evidence again peaks
at $N_{\rm knots} = 6$ and falls off after that.
However, this fall-off is slightly weaker than for the synthetic data
(Figure \ref{global:fig:FK_mock_logZ}). This could be an indication
that the signal underlying the EDGES low-band data is more complex
than the flattened Gaussian we used for the synthetic data. However,
this trend is only a weak argument, and in both cases the evidence peaks at
the same complexity level.
The ratio between the peak ($Z=\rm e^{267}$) and $N_{\rm knots} = 15$ ($Z=\rm e^{255}$) is still
very large ($>10^6$) so we can still safely truncate our analysis at that point.

\begin{figure*}
	\centering
	\includegraphics[width=0.8\linewidth]{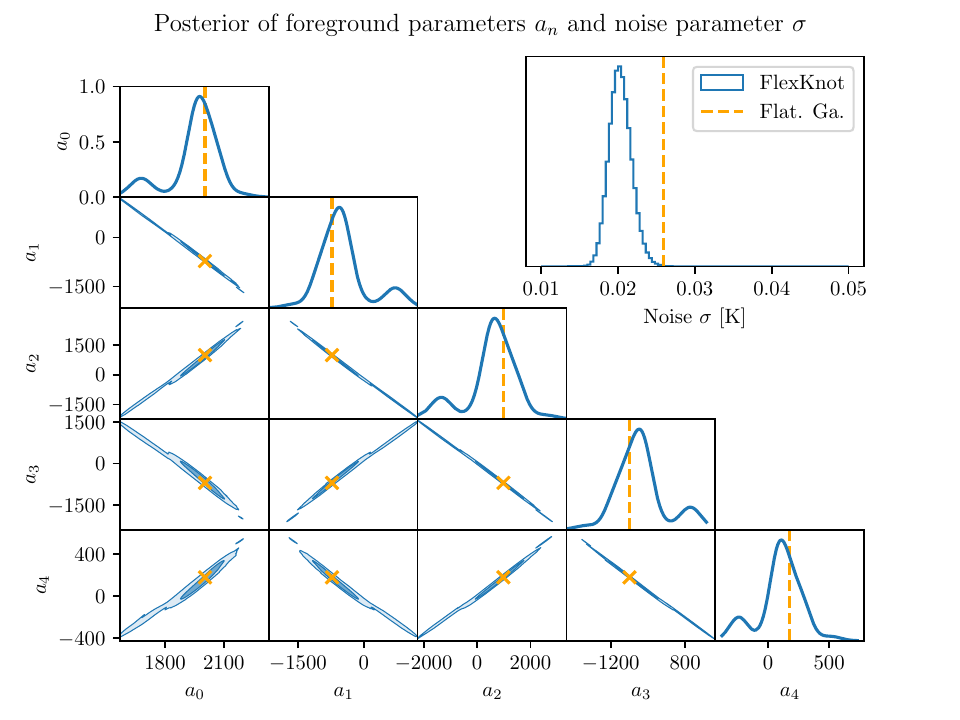}
	\caption{Noise and foreground parameter distribution for the FlexKnot
	signal fit to the EDGES low-band data. The foreground parameters $a_n$
	are degenerate as expected. We notice the FlexKnot fit is closer to the data and thus produces
    a lower noise posterior $\sigma_{\rm noise}$ (top right insert, compared
    to the dashed orange line).}
	\label{global:fig:FK_real_fgparams}
\end{figure*}

We show the posterior for foreground parameters in Figures
\ref{global:fig:FK_real_fgparams} ($a_n$) and
\ref{global:fig:FK_real_fgparams_tilde}
($\tilde a_n$). The foreground parameters $a_n$ values cover a large
range due to their degeneracy, while the transformed $\tilde a_n$ parameters
cover a narrower range.
The FlexKnot foreground posteriors cover a range near the
flattened Gaussian best-fit point (yellow crosses, best visible
in Figure \ref{global:fig:FK_real_fgparams_tilde}, lower panel) but do not
overlap as well as the synthetic data inputs did with the respective
posteriors (upper panel of Figure \ref{global:fig:FK_mock_fgparams_tilde}). This is
not unexpected as the EDGES data might not be well described by a
flattened Gaussian. The noise level of
$\sigma_\textrm{noise} = 0.020\,\mathrm{K} \pm 0.001\,\mathrm{K}$
\rr{is 
higher than the passive load test done
by the EDGES team \citep[0.015\,K,][]{Bowman2018}, but
closer to the expectation than } the flattened Gaussian fit
($\sigma_\textrm{noise} = 0.026\,\mathrm{K} \pm 0.002\,\mathrm{K}$)
showing that the FlexKnot functions allow a closer fit to the data.

\rr{Figure \ref{global:fig:FK_real_fgparams} also shows a
multi-modal distribution of foreground parameters. This is
expected, as the signal distribution (Figure \ref{global:fig:FK_real_lines})
is multi-modal and the foreground plus signal components must
add up to the data. We confirmed that the clusters seen in
Figure \ref{global:fig:FK_real_fgparams} are indeed the same
clusters as in Figure \ref{global:fig:FK_real_lines}.}

\section{Discussion}
\label{global:sec:discussion}
In this section, we want to discuss the strengths and limitations of the proposed methods, specifically the foreground orthogonalization we used
and its dependence on the chosen foreground model.

The foreground-orthogonal Gaussian Process fit is a new way to analyze
the \twocm global signal. It is a tool to extract and present
exactly the part of the data that cannot be explained by the foregrounds,
allowing us a direct comparison to theoretical predictions. This does
not assume that the theoretical signals are foreground-orthogonal,
we just apply a foreground-orthogonal projection to their theoretical description.
As discussed in section \ref{global:subsubsec:method_GP_orthogonal}, this projection is lossless, allowing us to obtain the likelihood
without the need for further sampling and marginalization.

Relying on the Gaussian Process fit to compare observations to theory models requires high confidence that the GP result indeed follows the data. By design, the GP foreground-orthogonal fit is a very robust method because the data always contains an unambiguous foreground-orthogonal component. We can verify this by plotting this component, i.e. the foreground-orthogonal projection of the data points themselves, along with the Gaussian Process posterior. We have done this in Figures \ref{global:fig:GP_mockresult} and \ref{global:fig:GP_realresult}, showing that the GP posterior contours indeed follow the data.

Finally, the foreground orthogonalization is based on distinguishing
between foreground-parallel and
foreground-orthogonal components of the signal.
Thus any result is dependent on the assumed foreground model.
This is not a unique problem to this method but
applies to all methods that rely on foreground modelling.
With the foreground model choice being an active topic of research
\citep[e.g.][]{Hills2018,Singh2019,Bevins2021maxsmooth},
we emphasize that our foreground orthogonalization can be applied directly
to any other foreground model with unbounded linear parameters,
such as equations (4), (7), (9) in \citet{Hills2018},
and equations (2) through (8) in \citet{Bevins2021maxsmooth}.
The only change required is to recompute the design matrix
$F_{ni}$ as described in section \ref{global:subsubsec:method_GP_orthogonal}.
We leave expanding the GP method to foreground models
with non-linear parameters or bounded parameter ranges for future work,
and note that the FlexKnot method can already be applied
to any parameterized foreground model
without the aforementioned restrictions.

\section{Conclusions}
\label{global:sec:conclusion}
In this paper, we have presented two model-independent approaches to explore
a \twocm measurement and find the signal (cosmological or systematic) that is present in addition to the foregrounds. The two methods achieve different and complementary goals.

Our first method, based on Gaussian Processes, can reliably extract the foreground-orthogonal component of the signal. This is an easy-to-implement and robust method that has been under-utilized in the literature so far.
We successfully apply the technique to both a synthetic data set and the EDGES low-band data \citep{Bowman2018}.
In the future, we expect this method to be used to compare observational data to theoretical signal predictions. In particular, theorists could explore which types of astrophysical models are especially \enquote{bright} in the foreground-orthogonal space and are thus easier to detect or rule out. Furthermore, one could approach investigating possible sources of systematics in the same way i.e. consider which kinds of systematics can explain the foreground-orthogonal part of the observed data.

Transitioning to our second method, we can summarize the gist of our first method as absorbing \textit{as much of the observed sky temperature as possible} into the foreground component, ending up with only the features that cannot be explained by the foreground.
Our second method on the other hand, based on FlexKnot, absorbs only as much of the observation into the foreground component \textit{as is beneficial to leave the simplest signal remainder}.

Our FlexKnot approach is the first free-form parameterization that can fit the entire \twocm signal present in the data (foreground-orthogonal and -parallel components) without relying on a theoretical model or pre-defined functional shape. Instead it utilizes a flexible interpolation scheme and 
Bayesian model selection to find the simplest signal that explains the data.
We successfully demonstrate its ability on a synthetic data set and present a novel analysis of the EDGES low-band data \citep{Bowman2018}. The multi-modal posterior distribution we find suggests that multiple signals are consistent with the data, comparable to individual results in the literature \citep[e.g.][]{Hills2018}.

For both of our methods, we briefly want to discuss the required instrument sensitivity. The Gaussian Process method relies on the foreground-orthogonal component of the data being significantly larger than the noise level (otherwise the Gaussian Process is consistent with zero), and the FlexKnot method indirectly relies on the same criterion (otherwise the simplest FlexKnot model to explain the data will be a constant). Thus we can only expect these methods to provide useful results if the foreground-orthogonal projection of the data is significantly larger than the noise level of the instrument. This is a similar sensitivity requirement to other methods, as a lower sensitivity would always allow a foreground-only fit to explain the data.

In the future, we expect the foreground-orthogonalization to be applied to a multitude of data sets, theoretical signal predictions, and different foreground models.
In particular, a combination with recent advances in foreground modelling \citep{Tauscher2018,Anstey2021,Shen2022} could provide the basis for future \twocm global signal analyses.

We are especially excited about extensions of the FlexKnot method. In particular, while distinguishing between cosmological and systematic contributions is not within the scope of this paper, there are properties we expect of the cosmological signal that can help us distinguish between the two. The FlexKnot method could be expanded to model the cosmological signal specifically \citep{Shen2023arXiv231114537S}, or to focus on certain types of systematics.

\section*{Acknowledgements}
We would like to thank Adam Ormondroyd, Harry Bevins, and Will Handley for
helpful discussions about our Nested Sampling procedure, and Thomas
Gessey-Jones for providing the simulation data used for Figure
\ref{global:fig:illustration_astrosignals}. We also thank the anonymous reviewer for their helpful feedback.

Besides the \texttt{PolyChord}, \texttt{anesthetic}, and \texttt{fgivenx}
libraries mentioned before, this work makes wide use of the
\texttt{NumPy} \citep{numpy2020}, \texttt{SciPy} \citet{2020SciPy-NMeth} \texttt{pandas} \citep{reback2020pandas}, and \texttt{Matplotlib} \citep{Hunter:2007} libraries.

SH acknowledges the support of STFC (grant code ST/T505985/1), via the award of a DTP Ph.D. studentship, and the Institute of Astronomy for a maintenance award.
HL was supported by the Institute of Astronomy as part of the
2021 summer internship programme.
FP was supported by the UK Medical Research Council programme MC\_UU\_00002/10, and the Engineering and Physical Sciences Research Council EP/R018561/1.
AF is supported by the Royal Society University Research Fellowship.

\section*{Data Availability}
The data underlying this article will be shared on reasonable request
to the corresponding author.

\bibliography{biblio_21cm}
\bibliographystyle{mnras}

\appendix

\section{Mean centering in GP fit}
\label{global:subsec:appendix_GPmeanzero}
\begin{figure}
	\centering
	\includegraphics[width=\linewidth]{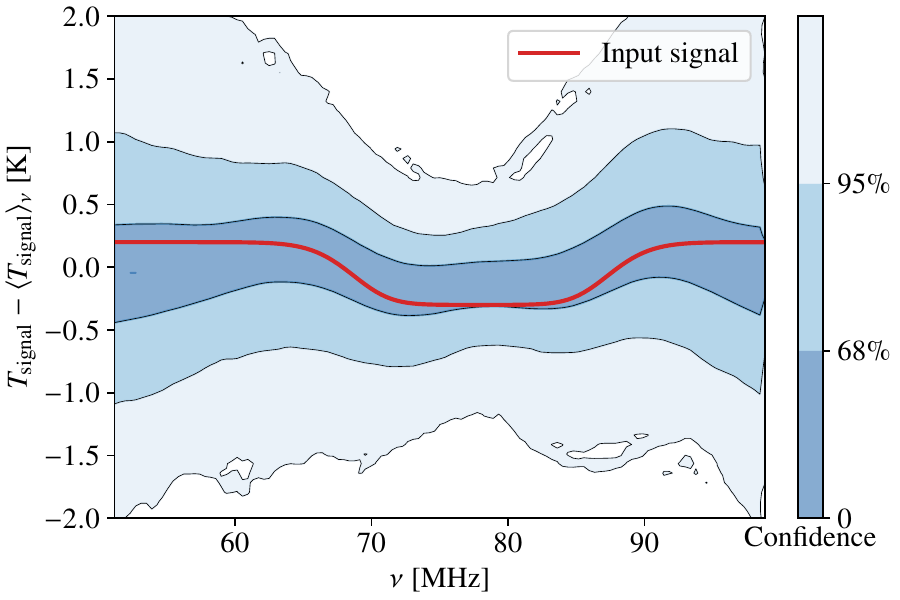}
	\includegraphics[width=\linewidth]{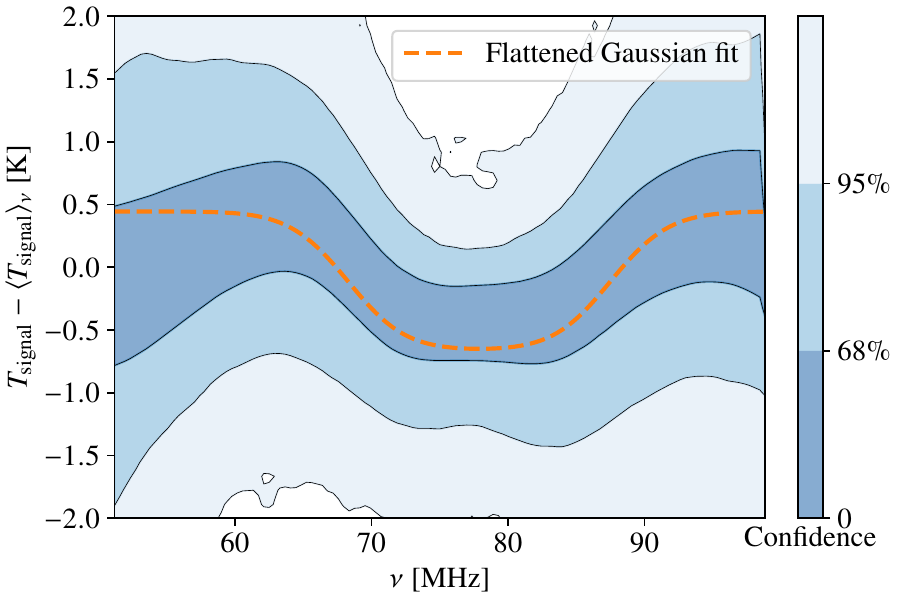}
	\caption{\rr{Gaussian Process posterior contours with mean-centering for the
	synthetic data (top) and EDGES low-band data (bottom). This Figure is the
	pendant to Figure \ref{global:fig:GP_nonorthog} but with the mean centering
	as described in section \ref{global:subsubsec:method_FK_meansubtraction}.
	The mean-centering helped reduce the degeneracy in the Gaussian Process
	contours somewhat, but does not solve the issue. The effect is not as
	strong as for the FlexKnot method (which is discussed in appendix 
	\ref{global:subsec:appendix_FKprojections}).}}
	\label{global:fig:appendix_GPmeanzero}
\end{figure}

\rr{In this section we test whether applying the mean subtraction
(section \ref{global:subsubsec:method_FK_meansubtraction}), that we apply to
the FlexKnot signals, can also remove the foreground degeneracy in the
Gaussian Process fit. Figure \ref{global:fig:appendix_GPmeanzero} shows the
GP posterior contours for the synthetic data (top) and EDGES low-band data
(bottom) with the mean subtraction applied.
As expected (see discussion in section \ref{global:subsubsec:method_FK_meansubtraction}), the
mean subtraction is not very effective and does not change the qualitative
result.
}

\section{Effect of signal prior choice in FlexKnot analysis}
\label{global:subsec:appendix_FKpriors}
In addition to the necessary priors in the knot parameters (where
we chose uniform priors from $[-2\,\mathrm{K}, 2\,\mathrm{K}]$
for the $T_{\rm signal}$ value of each knot) we can choose to add
an additional prior to encode what we expect of the signal.
In the main analysis, we chose a prior to limit the overall
contrast of the signal, that is the difference between the
highest and lowest signal value. We put a Gaussian prior
$\mathcal{N}(0\,\mathrm K, 1\,\mathrm{K})$ on this difference to
encode that we think a signal with a larger variation
is less plausible.

\begin{figure}
	\centering
	\includegraphics[width=\linewidth]{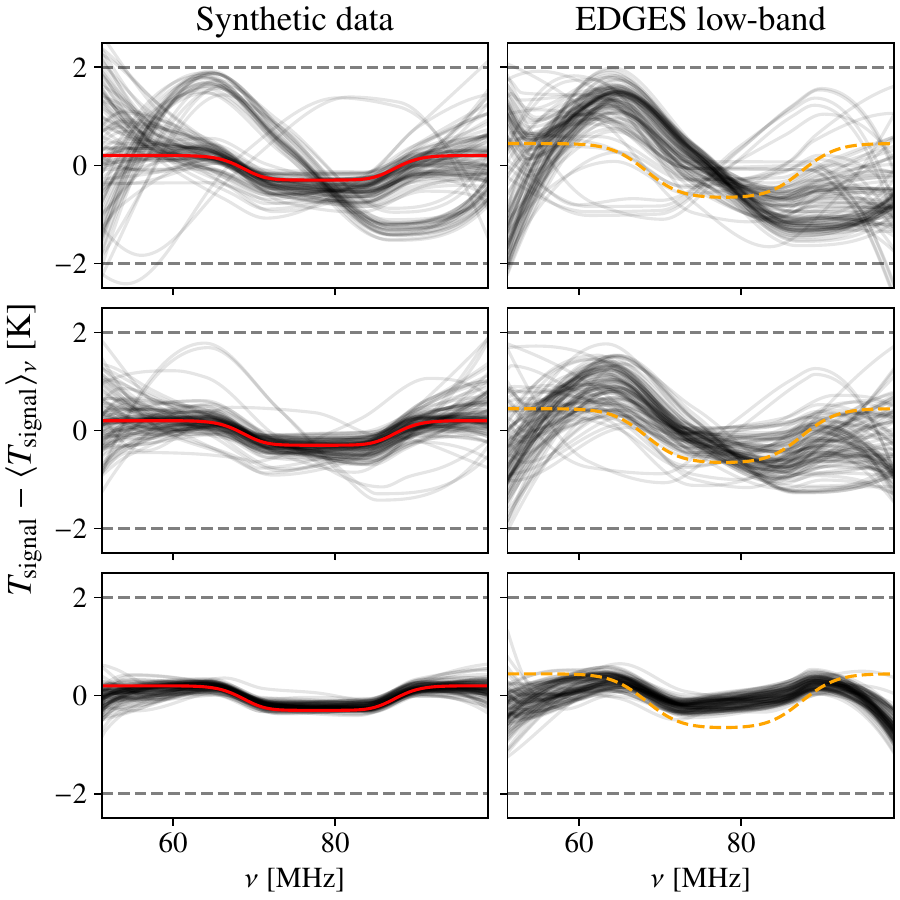}
	\caption{Comparison of prior choices for the FlexKnot signal fit. The first
	row shows runs with a uniform $[-2\,\mathrm{K}, 2\,\mathrm{K}]$ prior on
	the signal values, the second row shows runs with a Gaussian prior
	$\mathcal{N}(0\,\mathrm{K}, 1\,\mathrm{K})$ on the difference between the highest
	and lowest signal value, and the third row shows runs with a strong
	Gaussian prior $\mathcal{N}(0\,\mathrm{K}, 1\,\mathrm{K})$ on every
	signal value. We demonstrate the effect for both the synthetic data
	(left) and EDGES low-band data (right).
	We also add lines to indicate the input signal (red, left) or
	the flattened Gaussian best fit (orange, right) for easier comparison.}
	\label{global:fig:appendix_FKpriors}
\end{figure}

In Figure \ref{global:fig:appendix_FKpriors} we show the effect
of different such prior choices. The first
row shows just a uniform prior $[-2\,\mathrm{K}, 2\,\mathrm{K}]$
on the signal values. This is almost equivalent to no additional prior,
as the prior on the knot parameters approximately has the same effect
(but the signal could exceed the knot prior in the extrapolation). Note that
the plots show mean-subtracted signals which is why some lines reach
outside the $[-2\,\mathrm{K}, 2\,\mathrm{K}]$ range.
The second row shows the aforementioned Gaussian prior on
the signal contrast as we used in the main analysis.
The third row shows a Gaussian prior
$\mathcal{N}(0\,\mathrm{K}, 1\,\mathrm{K})$ on every value
$T_{\rm signal}(\nu_i)$ which we found to be too strong
and to suppress many posterior modes.

Note that in all cases we keep the uniform prior on the FlexKnot
coordinates $[-2\,\mathrm{K}, 2\,\mathrm{K}]$. We think this is appropriate
here since even a foreground-only fit produces residuals $\ll 2\,\mathrm{K}$,
but we emphasize that for other data a wider prior may need to be considered.

The left and right columns show the synthetic and EDGES low-band data respectively. We see that the prior choice overall mostly affects the weight between different posterior modes, as expected. For example, the synthetic data posteriors (left column) in the first row show a flattened Gaussian-shaped signal fitting the data, as well as a signal with a large $+2\,{\rm K}$ peak and $-1\,{\rm K}$ trough; the contrast prior in the second row gives more weight to the former and suppresses the latter more-extreme mode, as desired. The third row however is extremely constraining and artificially restricts the results to a single signal shape.

\section{FlexKnot mean-zero shift and foreground-orthogonal projection}
\label{global:subsec:appendix_FKprojections}
\begin{figure}
	\centering
	\includegraphics[width=\linewidth]{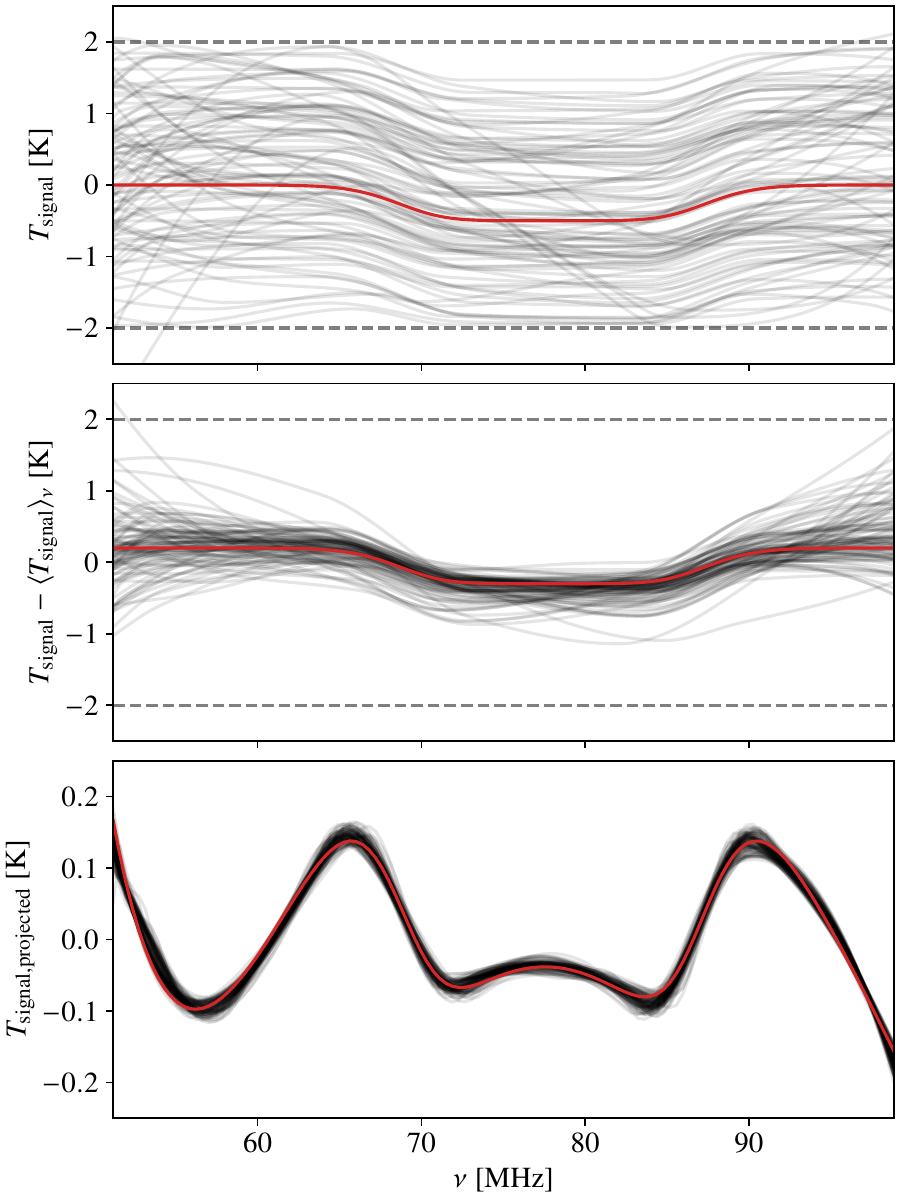}
	\caption{Different post-processing steps we consider for visualizing
	our FlexKnot results. \textbf{Top:} The raw FlexKnot posterior
	function samples. \textbf{Middle:} Posterior samples shifted to have
	zero mean, this is what we use in the main text.
	\textbf{Bottom:} Posterior samples but projected (in post-processing) into
	the foreground-orthogonal subspace, we just use this as a sanity check.
	All runs shown here use the synthetic data set and
	the Gaussian prior on signal contrast discussed in appendix \ref{global:subsec:appendix_FKpriors}.}
	\label{global:fig:appendix_meanzero_demo_DiffGaussian}
\end{figure}

When we obtain the FlexKnot parameter posterior we first have
a sample of knot parameters as illustrated in the lower panel of
Figure \ref{global:fig:illustration_knot_coordinates}. We then
turn each sample into the corresponding $T(\nu)$ function via
the FlexKnot interpolation. The resulting functions are shown
in the upper panel
of Figure \ref{global:fig:appendix_meanzero_demo_DiffGaussian}.
We can already read off the different function shapes, but due
to the trivial degeneracy of shifting the signal up and down
(this is not captured by the complexity penalty from the Bayesian evidence) these plots can be hard to interpret when there are many different shapes. Contour plots in particular are impossible to interpret in this format.

Thus, we remove this one degeneracy in post-processing by shifting all
signals to have zero mean, as shown in the middle panel of
Figure \ref{global:fig:appendix_meanzero_demo_DiffGaussian}.
This does not affect the functional shape as far as we are interested,
and makes the plots much more readable. This is the format we use for
all functional posterior plots in sections \ref{global:subsec:result_FK_mock}
and \ref{global:subsec:result_FK_real}, and appendix \ref{global:subsec:appendix_FKpriors}.

Finally, we could also project the FlexKnot signals into the foreground-orthogonal
subspace, as shown in the lower panel of Figure
\ref{global:fig:appendix_meanzero_demo_DiffGaussian} (note the different
$y$-axis). This is a useful check that the FlexKnot result is consistent
with the Gaussian Process result; all FlexKnot results are consistent
with the GP results in this projection.
We emphasize that the FlexKnot still describes
the full signal $T_{\rm signal}$ (and we apply the foreground-orthogonal
projection in post-processing just for this check) while the Gaussian Process
directly fits the foreground-orthogonal component $T_{\rm signal,\ projected}$.

\section{Transformed foreground parameter posteriors}
\label{global:subsec:appendix_fgparams}
In this section, we show the foreground parameter plots for the
transformed parameters $\tilde a_n$.
These parameters do not suffer from the degeneracies
of the non-transformed parameters $a_n$ and thus simplify
the comparison between the FlexKnot results and the
synthetic data inputs or flattened Gaussian best-fitting
parameters, respectively.

\begin{figure*}
	\centering
	\includegraphics{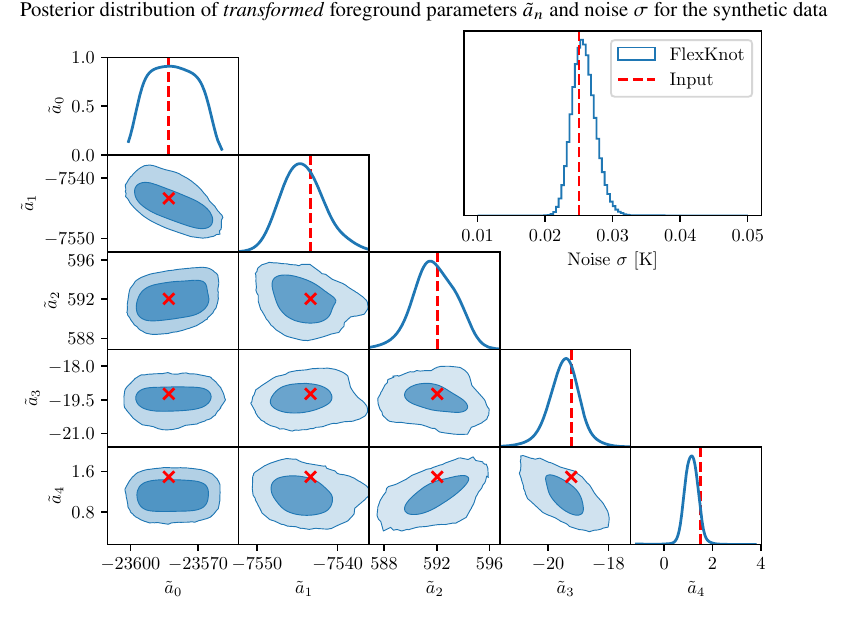}
	\par\vspace{8mm}
	\includegraphics{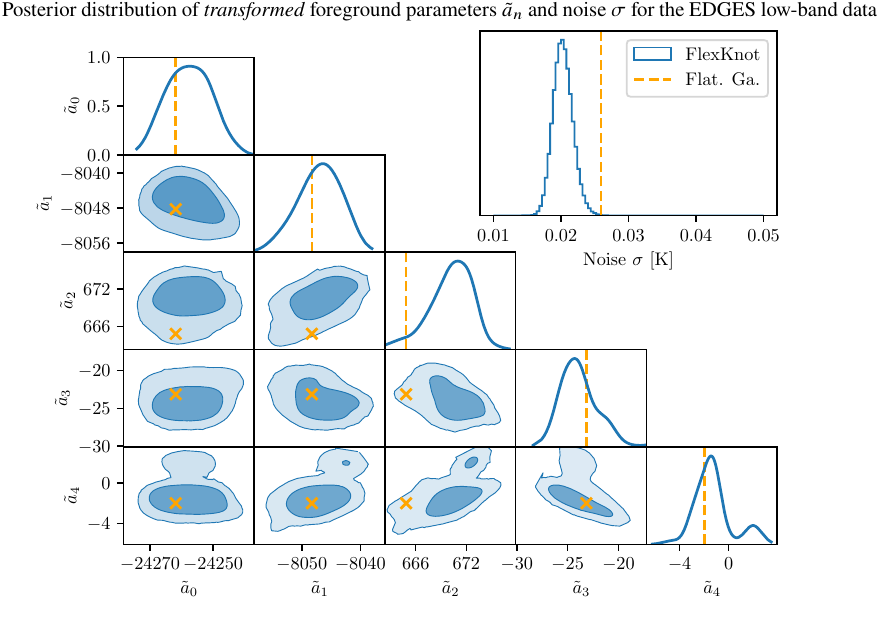}
	\caption{Posterior distributions of the transformed parameters
	$\tilde a_n$ (rather than the $a_n$ parameters shown in Figures
	\ref{global:fig:FK_mock_fgparams} and \ref{global:fig:FK_real_fgparams}).
	We show the FlexKnot foreground posterior distributions (blue contours) for the
	synthetic (\textbf{top}) and EDGES low-band (\textbf{bottom}) data sets.
	The top panel shows that the input parameters used for synthetic data
	(red markers) are well recovered by the FlexKnot fit.
	The bottom panel instead shows the foreground parameter values
	for the best-fitting flattened Gaussian model (orange markers).
	While these lie within the FlexKnot 95\% credibility region,
	they overlap less well with the FlexKnot posterior. This is expected
	because the FlexKnot method finds different signals than the
	flattened Gaussian fit.}
	\label{global:fig:FK_mock_fgparams_tilde}
	\label{global:fig:FK_real_fgparams_tilde}
\end{figure*}

In Figure \ref{global:fig:FK_mock_fgparams_tilde} we show the
synthetic data posterior (top panel), corresponding to Figure
\ref{global:fig:FK_mock_fgparams} in the main text, and
the EDGES low-band data posterior (bottom panel), corresponding
to Figure \ref{global:fig:FK_real_fgparams} in the main text.

\bsp	%
\label{lastpage}
\end{document}